# Dynamic Threshold Optimization – A New Approach?

*(Revised 05 June 2012)*

Richard A. Formato

Consulting Engineer &

Registered Patent Attorney

P.O. Box 1714, Harwich, MA 02645 USA

rf2@ieee.org

## Abstract

Dynamic Threshold Optimization (DTO) adaptively "compresses" the decision space (DS) in a global search and optimization problem by bounding the objective function *from below*. This approach is different from "shrinking" DS by reducing bounds on the decision variables. DTO is applied to Schwefel's Problem 2.26 in 2 and 30 dimensions with good results. DTO is universally applicable, and the author believes it may be a novel approach to global search and optimization.

## Problem Statement

In a bounded hyperspace $\Omega := \{ \vec{x} | x_i^{\min} \le x_i \le x_i^{\max}, \ i = 1,...,N_d \}$ (decision space), the $x_i$ being decision variables, determine the locations and values of the global maxima of the objective function $f(x_1, x_2, ..., x_{N_d})$, that is, max$\{ f(\vec{x}) : \vec{x} \in \Omega \subset \Re^{N_d}, f : \Omega \subset \Re^{N_d} \to \Re \}$. The value of $f(\vec{x})$ at each point $\vec{x}$ is its "fitness."

## Dynamic Threshold Optimization

DTO is conceptually quite simple. Figure 1 is a schematic illustration of how it works in a one-dimensional (1-D) DS. Objective function $f(x)$ is multimodal with many local maxima and a single global maximum, and the problem is to locate that maximum (coordinates and value). DTO bounds $f(x)$ from below using a series of successively increasing "thresholds," in effect compressing DS in the direction of the dependent variable (from "below") instead of, as is sometimes done, shrinking DS by reducing the independent variable's domain (from the "sides"). Locating the global maximum is easier in the compressed DS because unwanted local maxima are progressively filtered out as the "floor" (threshold) rises. Because DTO is a general geometric technique, it is algorithm-independent so that it can be used with any global search and optimization algorithm. Although DTO is described in the context of maximization, it can be applied to minimization with obvious modifications because $\max f(\vec{x}) = -\min f(\vec{x})$.

Procedure $OPT[q(\vec{x}), \vec{x}^*, q^*, q_{\min}]$ is a global search and optimization routine that returns (i) the $N_d$ coordinates $\vec{x}^*$ of a maximum of function $q(\vec{x})$, (ii) its value $q^*$, and (iii) a minimum value $q_{\min}$ (no coordinates). $OPT[\cdot]$ may comprise any search and optimization algorithm (singly or in combination with others) regardless of its type, deterministic, stochastic, or hybrid; and different algorithms may be used on successive calls to $OPT[\cdot]$.

DTO is initialized by applying $OPT[\cdot]$ to $f(\vec{x})$ without any threshold. Its return values then are used to define a starting threshold that subsequently is updated by applying $OPT[\cdot]$ to the auxiliary function $g(\vec{x}) = [f(\vec{x}) - T] \cdot U[f(\vec{x}) - T] + T$, where $T$ is the threshold value, and $U[\cdot]$ is the Unit Step function, $U(z) = \begin{cases} 1, & z \geq 0 \\ 0, & z < 0 \end{cases}$. Thus, for $f(\vec{x}) \geq T$, $g(\vec{x}) = f(\vec{x})$, whereas for $f(\vec{x}) < T$, $g(\vec{x}) = T$. On each successive pass, DTO *changes the topology* of the decision space being searched until a user-specified termination criterion is met (often maximum number of passes or fitness saturation). DTO pseudocode appears below:

***Algorithm DTO***

(i) **Initialization**

CALL $OPT[f(\vec{x}), \vec{x}_0^*, f_0^*, f_{\min}]$

SET $T_0$ (Starting Threshold – see text; typically $T_0 = f_{\min}$)

(ii) **Loop over successive thresholds**

$k \leftarrow 0$ (following standard notation $\leftarrow$ means "is set to")

$F^* = -N$ (initialize best overall fitness, very large number < 0)

DO UNTIL [Termination Criterion] (see text)

(a) $k \leftarrow k+1$ (increment pass #)

(b) CALL $OPT[g(\vec{x}), \vec{x}_k^*, g_k^*, g_{\min}]$ where

$g(\vec{x}) = [f(\vec{x}) - T_{k-1}] \cdot U[f(\vec{x}) - T_{k-1}] + T_{k-1}$

(c) IF $g_k^* >= F^*$ $\therefore$ $F^* = g_k^*, \vec{X}^* = \vec{x}_k^*$ where

$\vec{X}^*$ is the location of the best overall fitness

(d) UPDATE THRESHOLD: $T_k$ (see text)

LOOP

(iii) **Return:** $\vec{X}^*, F^* = f(\vec{X}^*)$ (best overall fitness: coordinates & value)

How to set DTO's starting threshold and how it is updated are determined by the algorithm designer. One obvious starting value is the minimum fitness returned by $OPT[\cdot]$, that is, $T_0 = f_{\min}$ (which seems to be the best default choice). But updating the thresholds $T_k$ as DTO progresses is more problematic because of the floor's profound impact on compressing DS. More and more local maxima are removed from DS as the threshold rises, so that effectively sampling DS becomes progressively more difficult (the landscape becomes flatter and flatter). In the limit of the floor rising to a global maximum, DS collapses to a plane, and there is no information available for performing a search. How well DS can be explored thus becomes more of an issue as the threshold rises, and the search algorithm's exploration characteristics become very important. One approach to setting $T_k$ is shown in Figure 1 in which successive thresholds are set to the best returned fitness, $T_k = g_k^*$, but this approach has not worked well in numerical tests because a good search algorithm often sets the threshold too high too early in the run. The 2-D example that follows employs a different approach, and it clearly illustrates the effect of flattening DS too much.

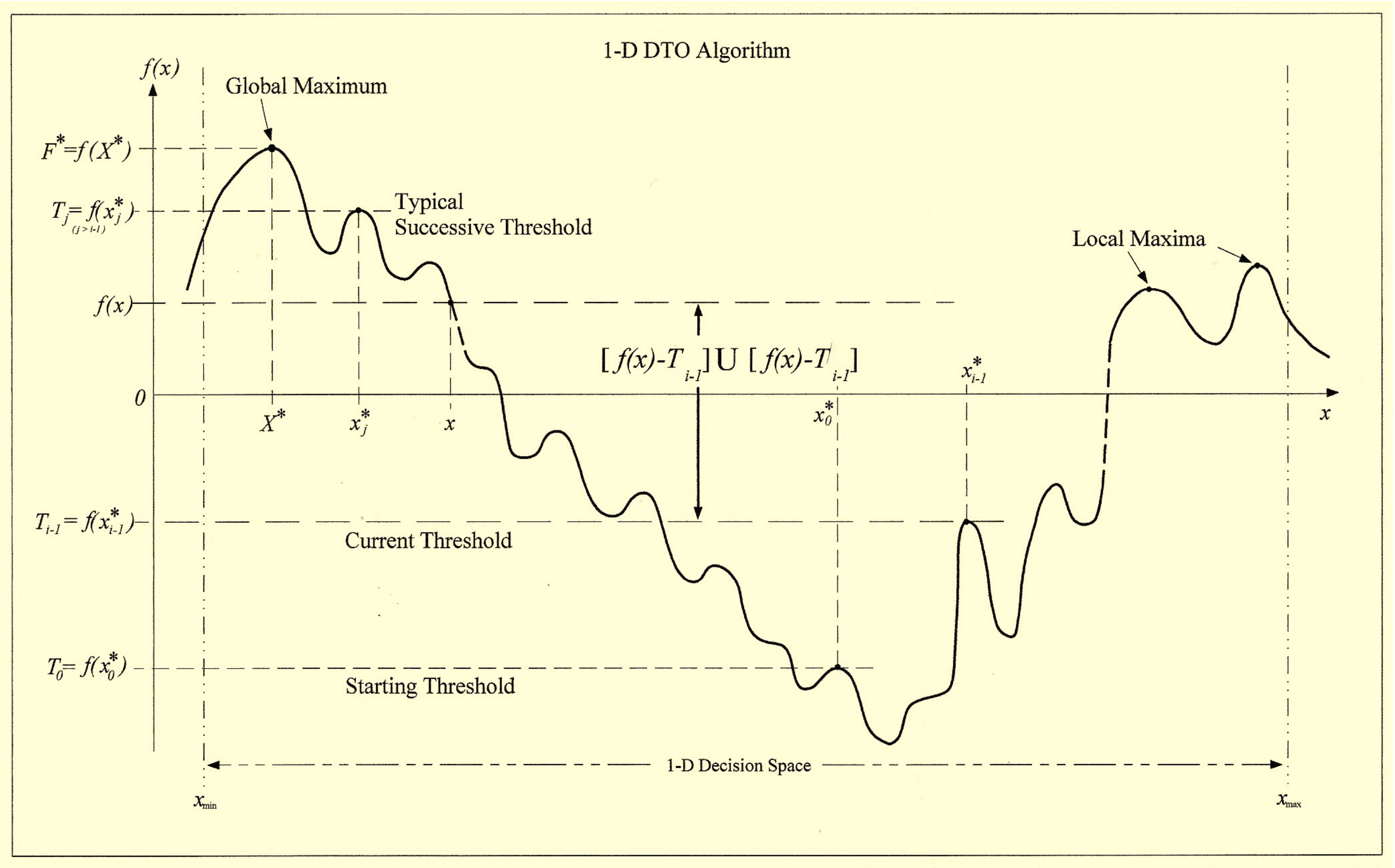

**Figure 1. DTO Concept with Thresholds at Successive Local Maxima.**

As an example of how it works, DTO was applied to Schwefel's Problem 2.26 in 2-D using Central Force Optimization (CFO). In an $N_d$-dimensional decision space, this objective function is $f(x)=\sum_{i=1}^{N_d}[x_i \sin(\sqrt{|x_i|})]$, $-500 \le x_i \le 500$. It has a single global maximum of $418.9829 \cdot N_d$ at $[420.9687]^{N_d}$ ([1] @ p.4467). The 2-D global maximum is $837.9658$ @ $(420.9687, 420.9687)$. DTO was implemented with a number of passes $P=10$ with a progressively increasing threshold computed as $T_k = F_{\min} + \frac{C_{th} k}{P}(F^* - F_{\min})$, $k=1,\ldots,P$ (no threshold applied for $k=0$), where $F^*$ and $F_{\min}$, respectively, are the best and worst overall fitnesses returned through pass $k$. The coefficient $C_{th}=0.98$ in this case is included to keep the threshold far enough below the global maximum that the DS is not compressed into a plane. This formula for setting the threshold was chosen as much for its ability to illustrate the DTO concept (see plots below) as for its ability to produce good results, and there no doubt are countless other approaches to setting the threshold that will work better.

The number of CFO probes was initialized to $N_p=4$, and it was doubled on each successive pass in order to enhance CFO's exploration. Each run comprised $N_t=25$ time steps. While CFO is an inherently deterministic search and optimization metaheuristic, in this case it was implemented with a random initial probe distribution (IPD) instead of the usual "Probe Line" IPD [2]. The reason for this change, again, was to enhance

CFO's exploration in the progressively flatter DS. The complete Power Basic source code listing appears in the Appendix.

The DTO/CFO algorithm returned a best overall fitness of $F^* = 837.965574726692$ at the point $(x_1, x_2) = (421.007498176246, 420.959700549993)$ using a total of 106,392 function calls. The errors in the fitness and in the coordinates, respectively, are 0.0002253 and (-0.0387982,0.0089995) [computed as known-DTO], which are quite small. Table 1 summarizes the DTO threshold evolution pass-by-pass and CFO's best fitness.

**Table 1. DTO/CFO Results for 2-D Schwefel Problem 2.26**

| ***Pass #*** | ***Threshold*** | ***Best Fitness*** |
|---|---|---|
| 1 | *none* | 580.2973878 |
| 2 | -347.955 | 837.8781823 |
| 3 | -196.617 | 719.2845548 |
| 4 | -70.522 | 837.8956233 |
| 5 | 55.580 | 837.9282076 |
| 6 | 181.693 | 837.9654027 |
| 7 | 307.815 | 837.9504301 |
| 8 | 433.919 | 837.9647313 |
| 9 | 560.022 | 837.9650286 |
| 10 | 686.126 | 837.9655747 |

Figure 2 shows how DTO compresses DS as its threshold increases. The objective function is plotted at each of the 10 passes. The first pass (no threshold) shows the Schwefel Problem 2.26's complex landscape. It is highly multimodal with many similar amplitude local maxima. As DTO progresses more and more of these maxima are filtered out because the DS floor is higher and higher relative to the single global maximum. At pass #8, for example, 16 local maxima are visible, whereas at thresholds #9 and 10, respectively, the number of maxima falls to 8 and to 3. On the last pass the global maximum is clearly visible on the right side of the plot.

**Figure 2. DTO Compression of 2-D Schwefel Problem 2.26 DS using Successive Thresholds**

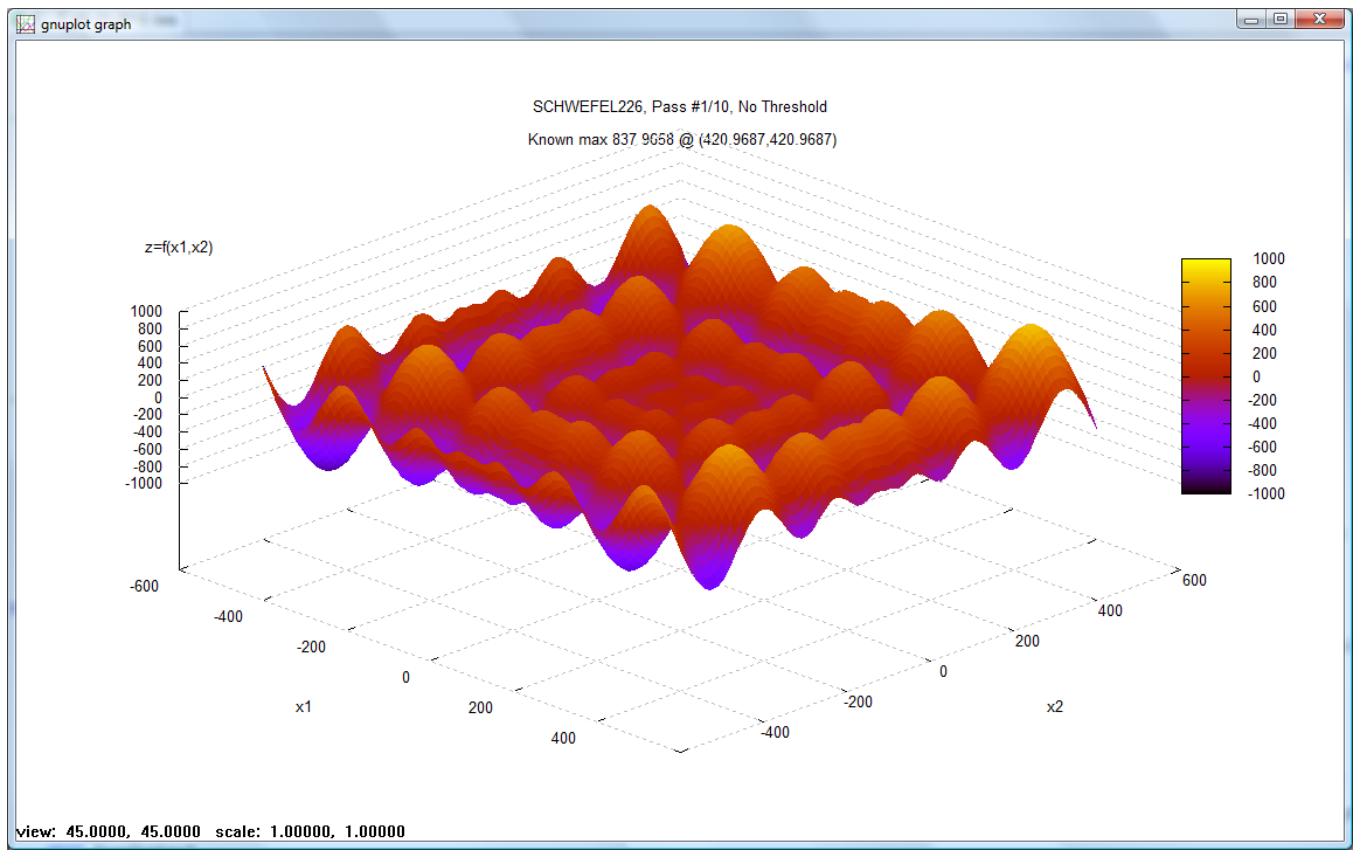

Pass #1

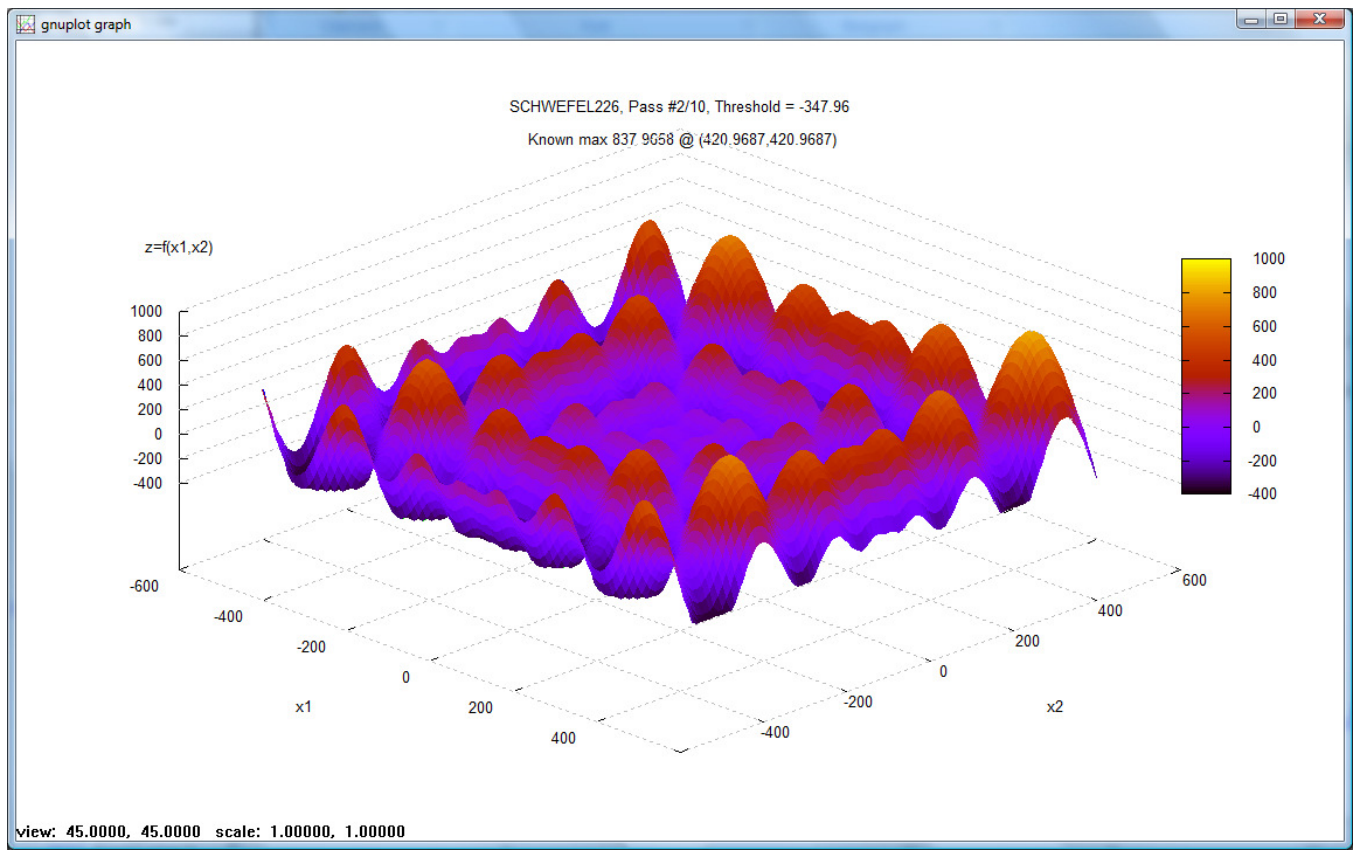

Pass #2

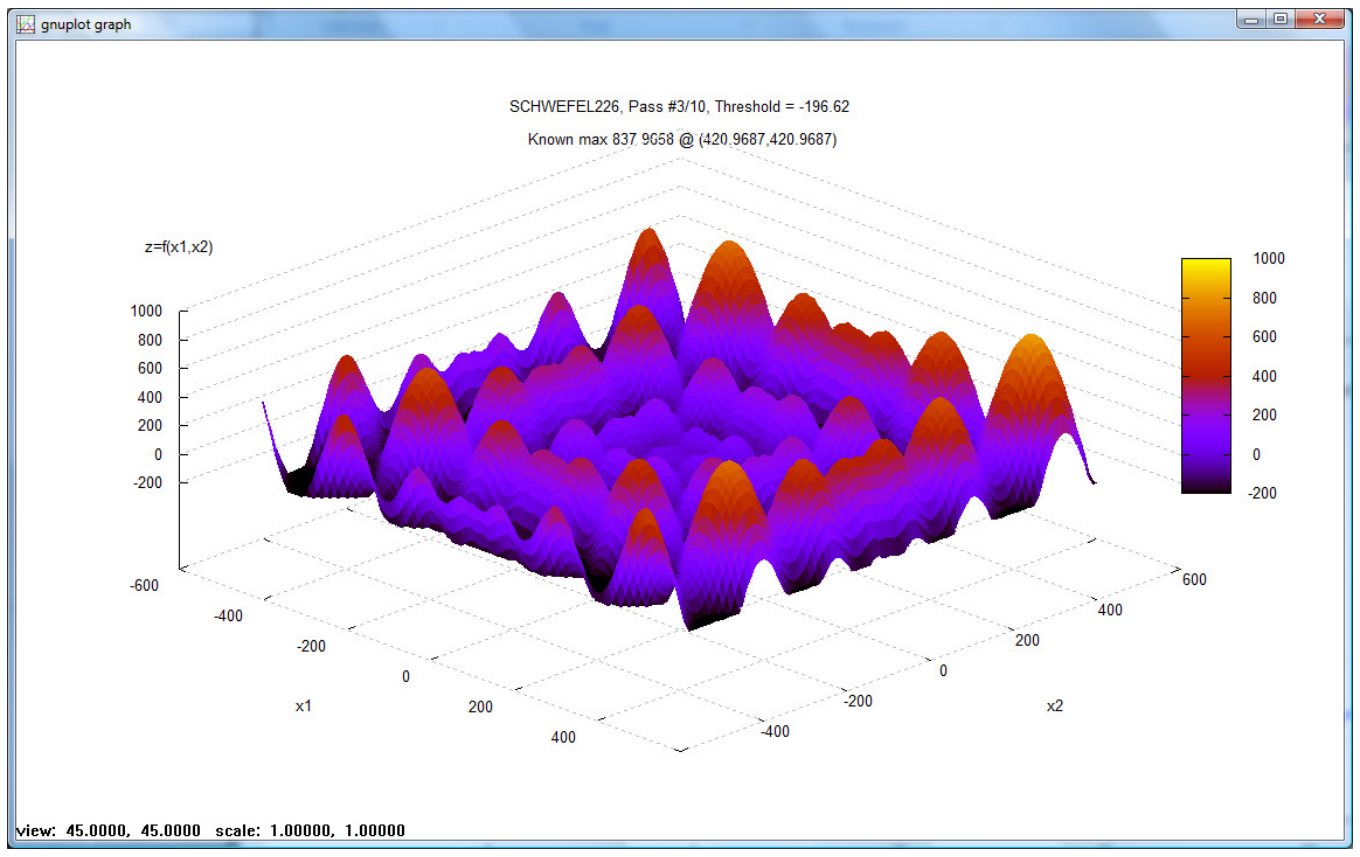

Pass #3

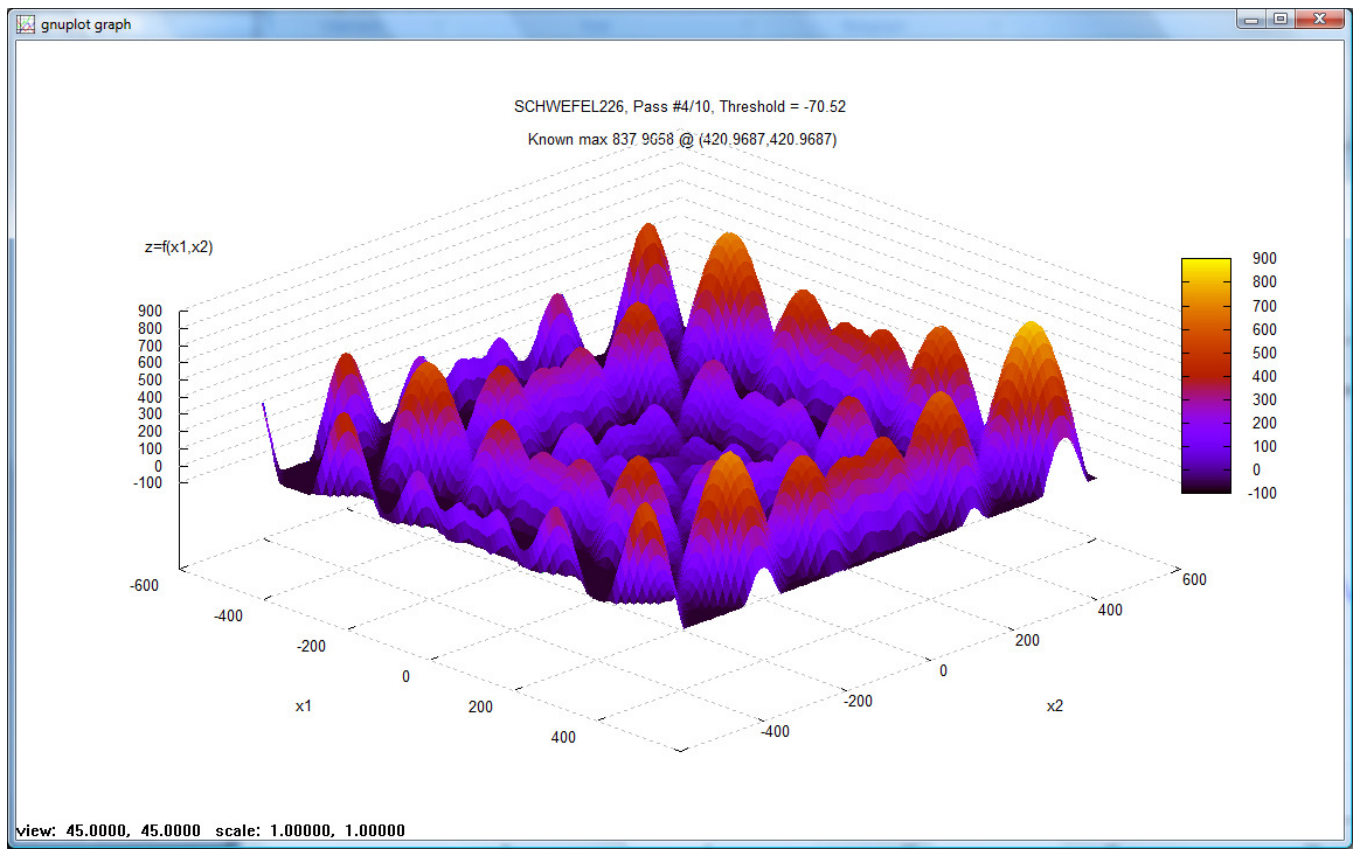

Pass #4

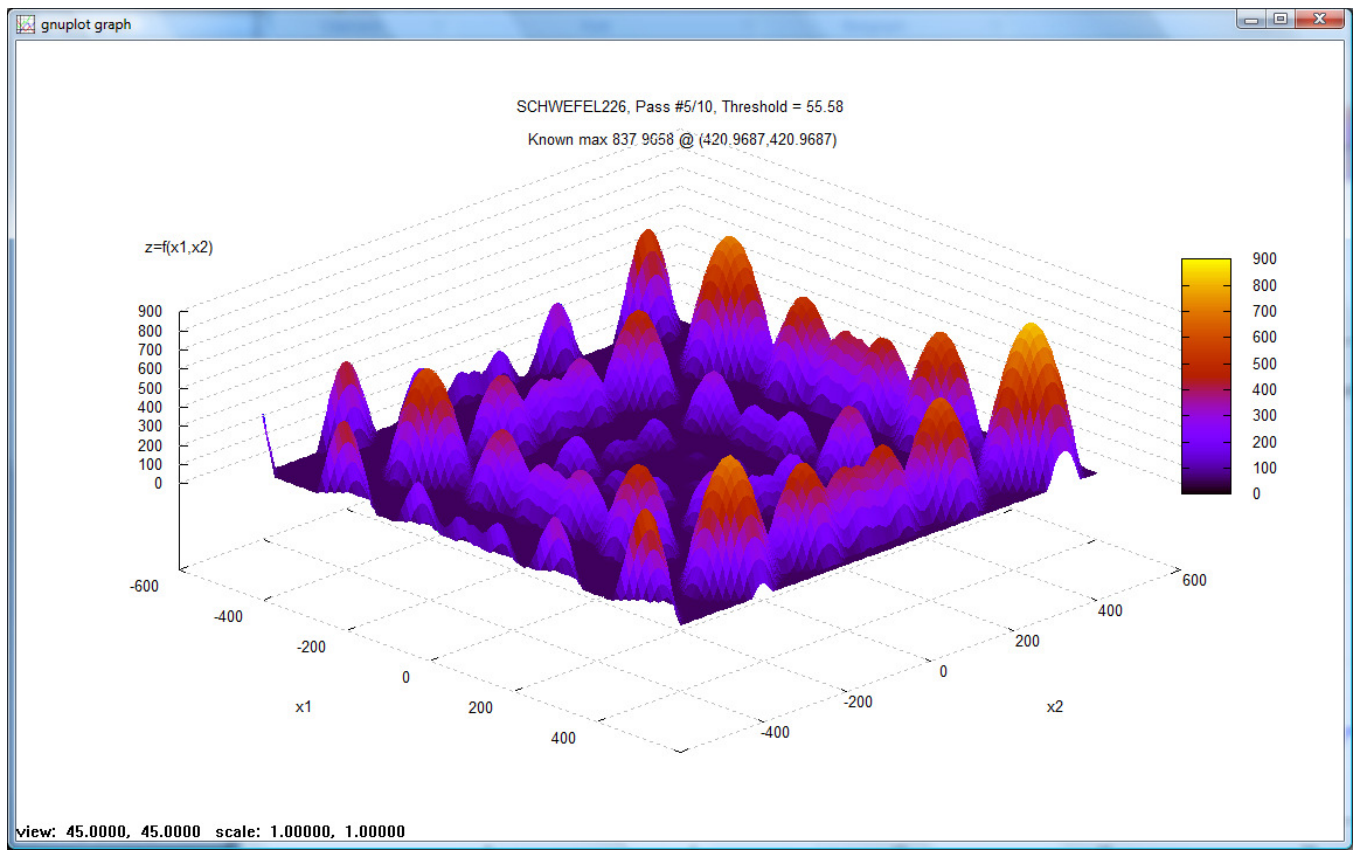

Pass #5

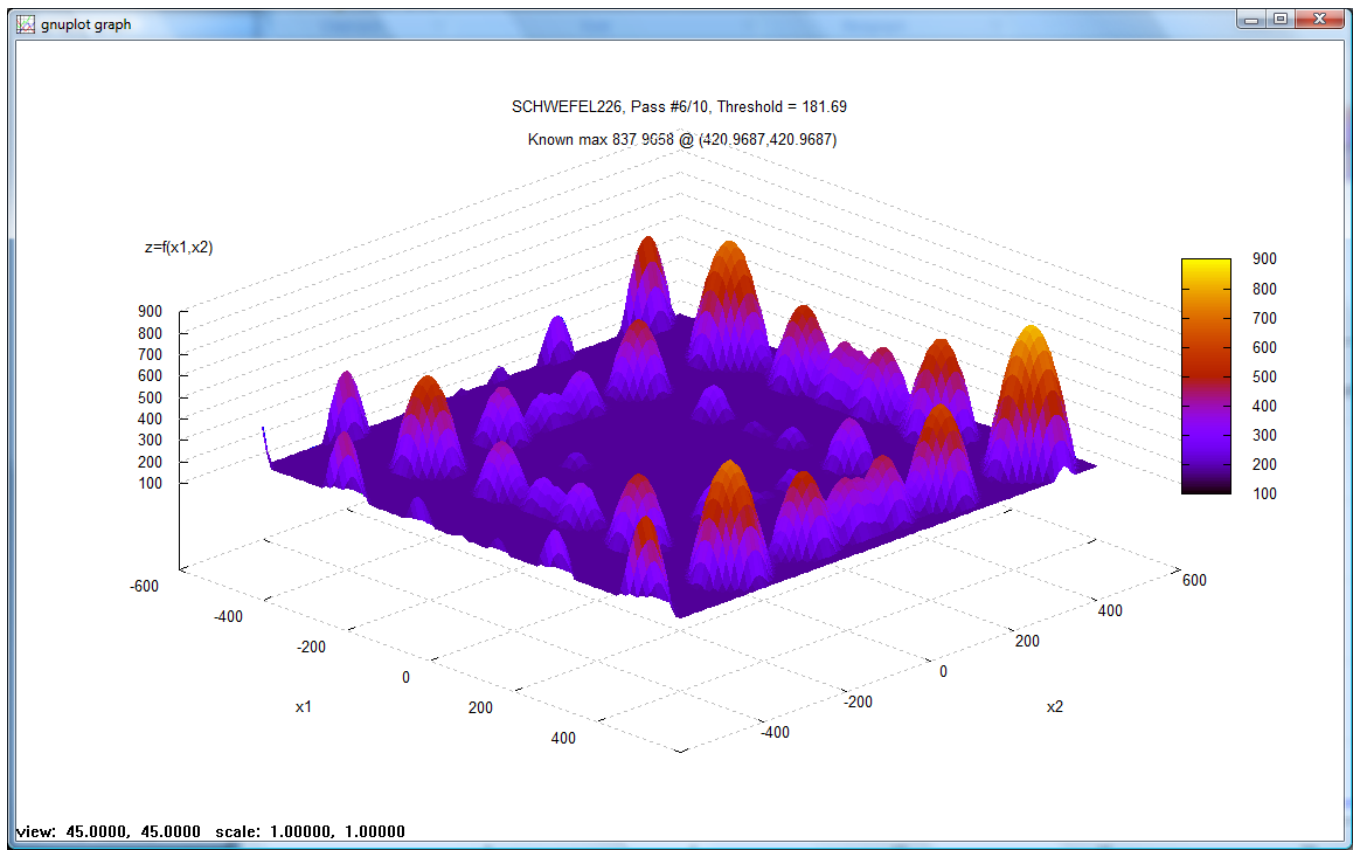

Pass #6

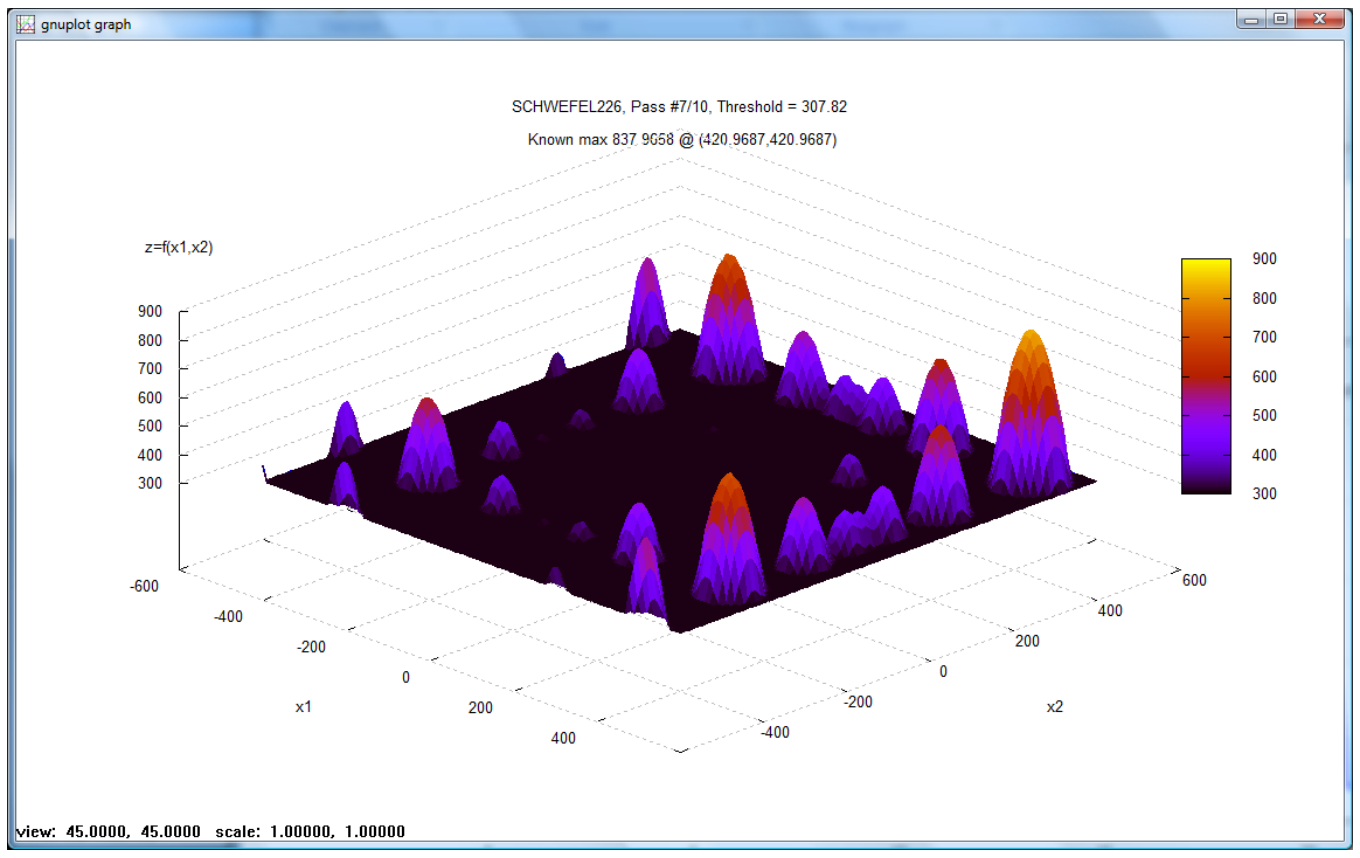

Pass #7

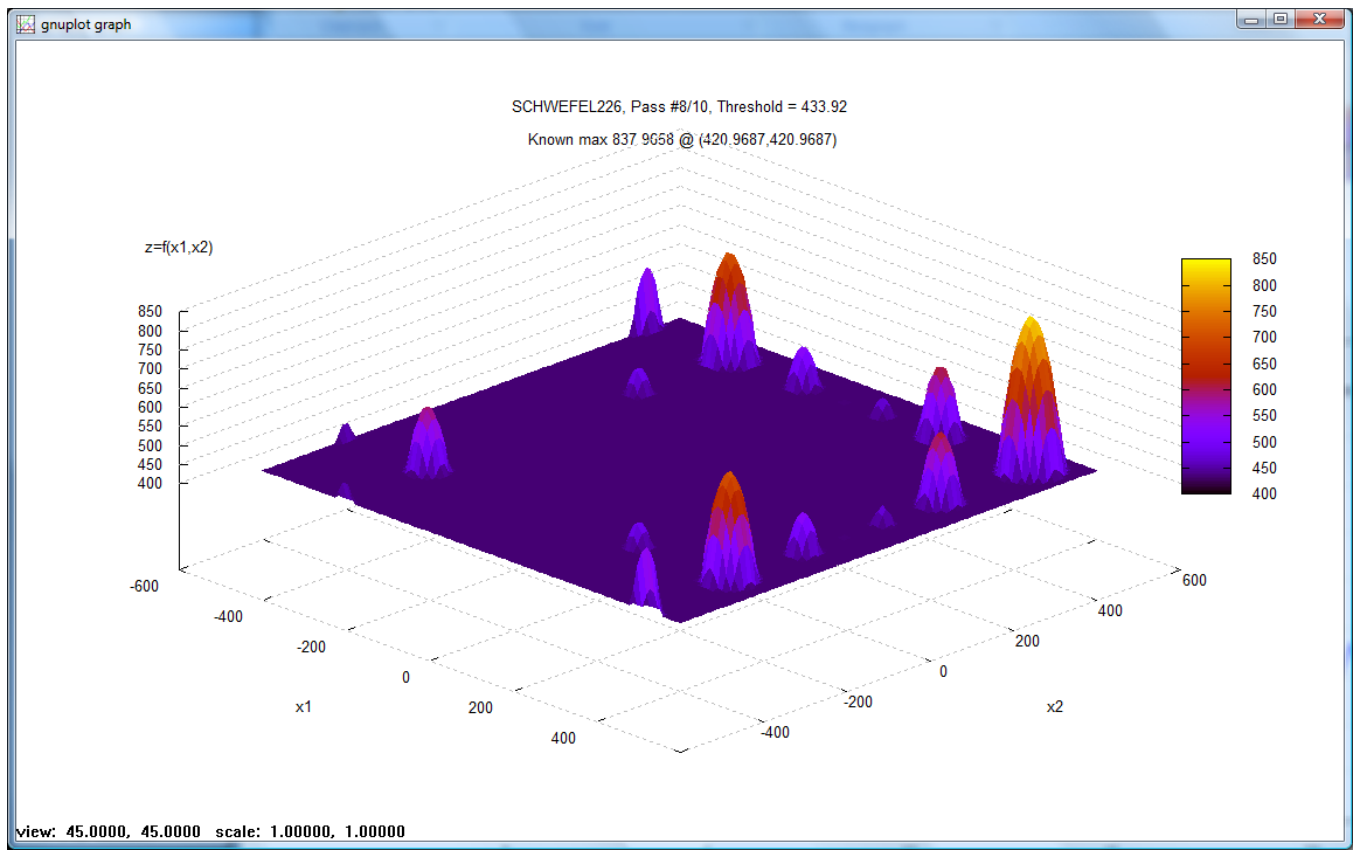

Pass #8

*

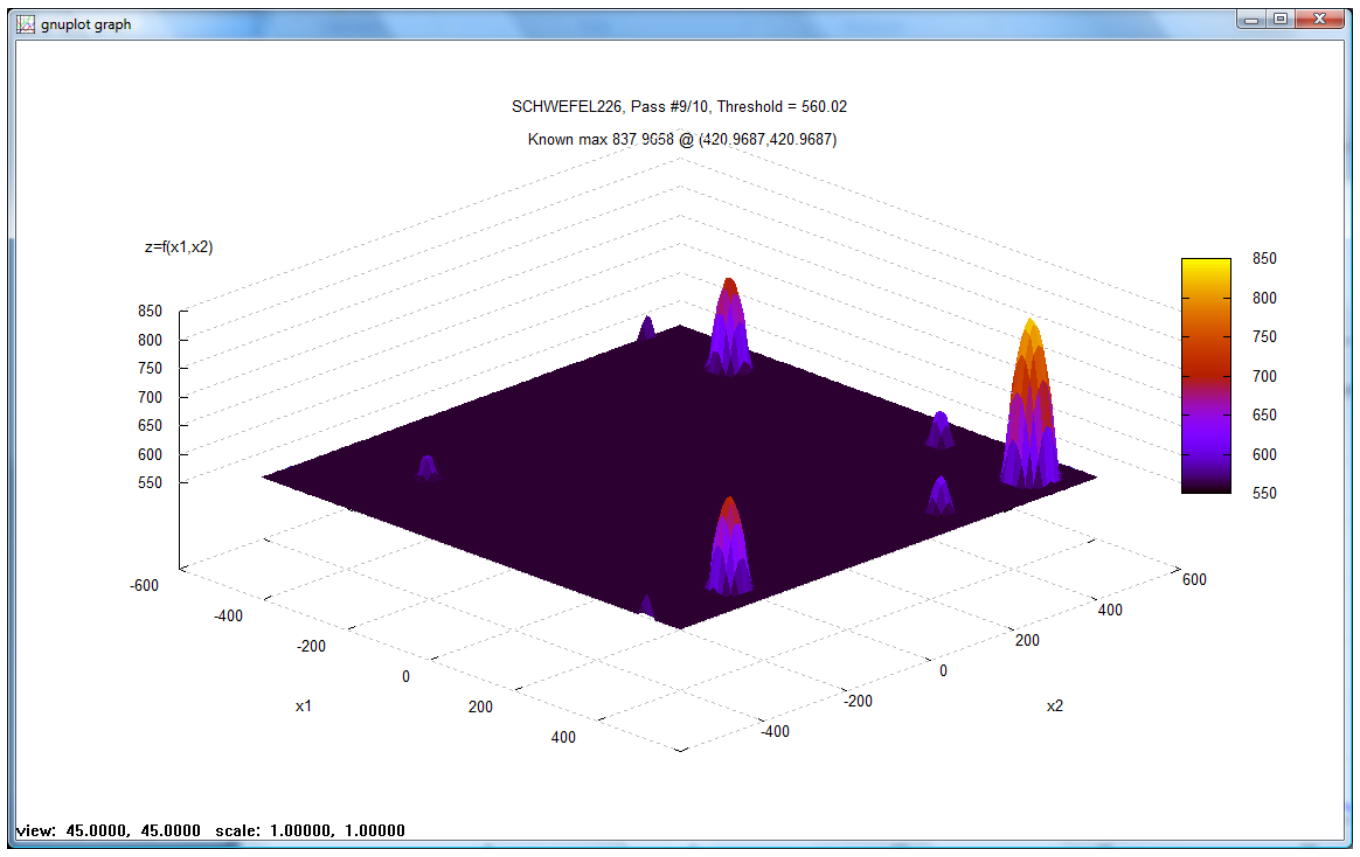

Pass #9

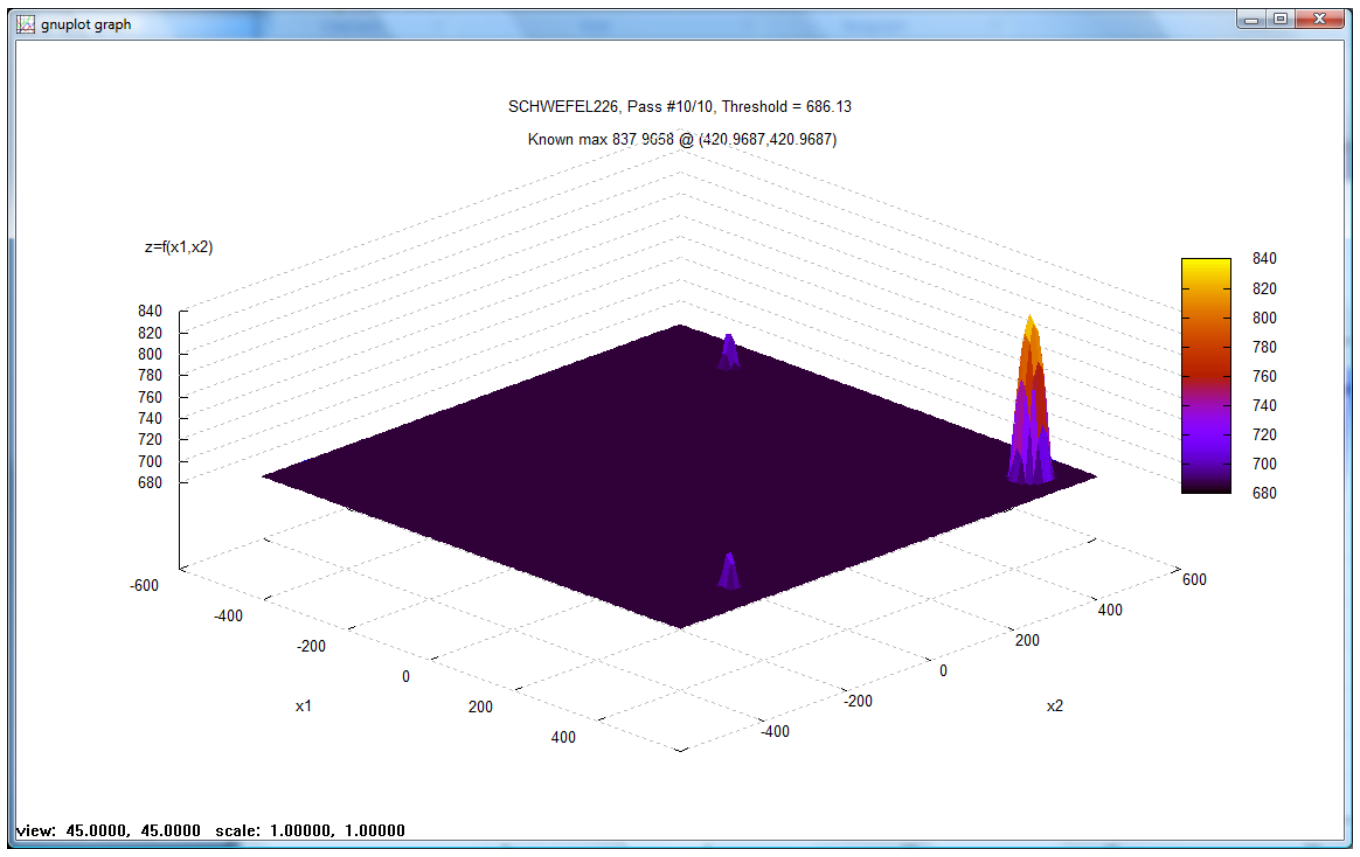

Pass #10

DTO also was tested against Schwefel 2.26 in 30-D. Six passes were made using the linear threshold scheme described above, but with $C_{th} = 0.6$. Unlike the 2-D case, a deterministic CFO implementation was used with a Probe Line IPD and $\gamma = [0,1], \Delta\gamma = 0.1$ (see [2] for details). Other CFO parameters were the same as above except for $N_t = 15$. Passes 2 through 6, in order, had calculated thresholds of $-10{,}176.09$, $-7{,}828.153$, $-5{,}480.216$, $-3{,}132.279$, and $-212.722$. The best fitness returned by DTO was $12{,}569.28$ at the point $x_i = 420.7353, i = 1,..,29$, $x_{30} = 420.7662$. This result is quite good compared to the known maximum of $12{,}569.487$ (fractional error $1.647\text{x}10^{-5}$) requiring a total of $44{,}352$ function evaluations.

## Speculation

DTO is optimization algorithm-independent because it is fundamentally geometrical in nature. Procedure $OPT[q(\vec{x}), \vec{x}^*, q^*, q_{\min}]$ can be *any* global search and optimization routine, some combination of routines, different ones on successive DTO passes, or perhaps *none at all*. This observation raises the possibility of a new optimization approach that does not rely, as typically is the case, on a metaheuristic based on a metaphor drawn from Nature, or, for that matter, on any existing optimization methodology, heuristic or otherwise. Instead, it may be possible to develop a new optimization algorithm using only DTO's geometrical approach.

One possible approach might be to implement $OPT[q(\vec{x}), \vec{x}^*, q^*, q_{\min}]$ as a group of quasirandom (QR) samplings of DS at each DTO threshold (any sampling scheme can be used, but QR is attractive because these sequences are deterministic). This approach is especially attractive because of its simplicity. The data in each group could be used to develop statistics characterizing DS's topology at that threshold. Those statistics, in turn, can provide a measure of the likelihood of locating maxima. As DTO's threshold moves up, any peak at or below the floor cannot be a global maximum (unless DS is compressed into a plane). As DS is progressively compressed, QR sampling will return more and more sample points on the floor, that is, points at which there is no maximum of any kind. Repeatedly sampling a given threshold develops a picture of where the current maxima (local and global) can be located. In the limit, every point on the floor would be visited, and the global maxima located precisely. Of course, only a finite number of runs can be made, but it seems likely that very good statistics could be developed fairly quickly as DTO's threshold increases. At a minimum, this approach should be able to provide a reliable estimate of the likelihood of locating global maxima.

Brian Hayes' excellent article on QR sequences [3] may provide a blueprint for a DTO-QR optimization algorithm. For example, the problem of determining the area of a leaf is analogous to computing the area under $f(\vec{x})$'s maxima (peaks) projected onto a particular DTO threshold. If the compressed DS is sampled (QR or otherwise), $1-\frac{N_{th}}{N_s}$ estimates the probability of being within the peaks' projections ($N_{th}$ and $N_s$ being the number of sampling points falling *on* the threshold and the *total* number of points, respectively). Repeating this procedure *on each threshold* a sufficient number of times builds confidence in the estimate. Contrary to what might be intuitive, the objective of this approach is to reduce this probability to zero as the threshold increases. Zero probability of being within the maxima's projections corresponds to the threshold being at a global maximum because DS has been compressed onto a plane. If this happens, as pointed out above, all information on the maximum's location is lost, so as DTO progresses information on where maxima are found must be preserved in order to determine the global maximum's coordinates.

Besides changing DS's topology from below, statistics gathered as described above may be useful in shrinking DS from the "sides," that is, truncating $f(\vec{x})$'s domain of definition to create a smaller search space that is more easily explored. This might be accomplished by grouping proximate sample points above the threshold, that is, points within the "footprints" (projections) of local maxima, and then breaking the domain into smaller regions containing each footprint. The likelihood of locating all footprints on a given threshold increases with the number of $OPT[\cdot]$ runs made at that threshold.

Of course, all of these remarks are pure speculation at this point. Whether or not implementing some of these ideas may lead to a new and effective optimization methodology is an open question. But DTO appears to hold enough promise to be investigated further. One approach might be to initialize DTO with a deterministic algorithm such as CFO with a Probe Line IPD, because it tends to converge quickly to the vicinity of global maxima, followed by QR-based exploration as described above (or possibly a stochastic algorithm) because of potentially improved exploration.

## Conclusion

DTO appears to be an effective technique for adaptively changing the topology of the decision space in a multidimensional search and optimization problem. DTO should be useful with any search and optimization algorithm. Bounding DS from below removes local maxima, and as the threshold or "floor" is increased, more and more local maxima are eliminated. In the limit, DS collapses to a plane whose value ("height") corresponds to the value of the global maximum. In that case, DS contains no information as to the global maximum's location, but the maximum's value is known precisely. In order to preserve location information, the DTO threshold should not be set too high, thereby retaining enough structure for efficient DS exploration.

There are many unanswered questions concerning how DTO should be implemented. For example, there almost certainly are better ways to set the threshold than the simple linear scheme used here. Thresholds that are progressively closer together probably will work better. Another question arises in connection with what optimization algorithm should be used. Even though DTO is algorithm-independent, it may work best when different algorithms are combined to take advantage of their different strengths and weaknesses. For example, CFO, which is inherently deterministic, often converges very quickly to the vicinity of a global maximum (good exploitation). But its very determinism inhibits exploration in decision spaces with "sparse" structure (mostly planar, few local maxima). By contrast, stochastic algorithms (for example, PSO, ACO, or DE) exhibit better exploration, but they completely lack repeatability when implemented using the true random variables in their underlying equations (computed from probability distributions). Combining a deterministic algorithm used first with a stochastic one used later may provide better results by emphasizing exploitation early in the run and exploration later in the run. Or, in the case of CFO, it might be started deterministically and then switched to stochastic mode (recall that the CFO used here was stochastic for the 2-D Schwefel 2.26 and deterministic for the 30-D case). Another improvement might utilize "lateral" DS compression on one of DTO's thresholds. It may be possible in the DTO-compressed DS to reliably determine the global maximum's approximate location and, based on that information, shrink DS "from the sides" or "laterally" (reduce the domain of definition), making it easier to search the smaller DS. If DTO is a novel approach to optimization, as the author believes it is, then all of these possibilities merit consideration as fruitful areas of research, and the author hopes that this note will encourage such work.

The source code listing in the Appendix is available in electronic form upon request to the author; please email requests to `rf2@ieee.org`.

***2 June 2012***
***Revised 5 June 2012 (typos & reference [1] corrected; material added)***
***Brewster, Massachusetts, USA***

## References

[1] Wang, J., "Particle Swarm Optimization with Adaptive Parameter Control and Opposition" *J. Computational Information Systems*, Vol. 7, No. 12, 2011, pp. 4463-4470 (http://www.jofcis.com).

[2] Formato, R. A., "Parameter-Free Deterministic Global Search with Simplified Central Force Optimization," in **Advanced Intelligent Computing Theories and Applications** *(ICIC2010)*, Lecture Notes in Computer Science (D.-S. Huang, Z. Zhao, V. Bevilacqua, and J. C. Figueroa, Eds.), LNCS 6215, pp. 309–318, Springer-Verlag Berlin Heidelberg, 2010.

[3] Hayes, B., "Quasirandom Ramblings," *American Scientist* magazine, vol. 99, July-August 2011, pp. 282-287 (www.americanscientist.org).

## Appendix - DTO Source Code Listing

NOTE: This listing implements deterministic CFO used for the 30-D Schwefel Problem 2.26. The source code listing for the pseudorandom version of CFO used for the 2-D Schwefel is in the original version of this note (dated 2 June 2012).

```
'Program 'DTO_TEST_06-05-2012.BAS' compiled with
'Power Basic/Windows Compiler IDE 10.03.0102 (www.PowerBasic.com).

'THIS PROGRAM IMPLEMENTS DTO (DYNAMIC THRESHOLD OPTIMIZATION) USING A
'CENTRAL FORCE OPTIMIZATION (CFO) "CORE."  CFO IS AN INHERENTLY DETERMIN-
'ISTIC ALGORITHM.  THE USUAL "PROBE LINE" IPD IS USED TO ILLUSTRATE DTO
'USING A DETERMINISTIC ALGORITHM.  CFO OFTEN IS IMPLEMENTED WITH
'DECISION SPACE "SHRINKING" TO IMPROVE CONVERGENCE, BUT THAT FEATURE
'IS NOT USED IN THIS DTO VERSION.  SEE DISCUSSION IN THE ACCOMPANYING
'ARTICLE FOR ADDITIONAL DETAILS.
'===========================================================================

'LAST MOD 06/05/2012 ~1917 HRS EDT

'ADDED 30-D SCHWEFEL EXAMPLE 06-05-2012

'THIS SAMPLE DTO/CFO PROGRAM CONTAINS THREE 2D TEST FUNCTIONS: RASTRIGIN, SGO,
'AND SCHWEFEL226 (SELECTED BY SETTING VARIABLE 'UseFunction$').

'ON-SCREEN PLOTTING INCLUDED; REQUIRES THE SEPARATE PROGRAM 'wgnuplot.exe'.
'SEE http://www.gnuplot.info/ FOR DETAILS.

'********************************************************************************

'(c) 2012 Richard A. Formato.  All Rights Reserved Worldwide.

'********************************************************************************
'THIS PROGRAM IS FREEWARE AND MAY BE COPIED AND DISTRIBUTED WITHOUT LIMITATION AS
```

```
'LONG AS THERE IS NO CHARGE OR FEE OF ANY KIND, INCLUDING "BUNDLING" OR "TIE-IN"
'CHARGES OR FEES OR ANY OTHER TYPE OF COST.
'*********************************************************************************

#COMPILE EXE

#DIM ALL

DEFEXT A-Z

'--------------------------------------

GLOBAL MinDistanceAboveThreshold AS EXT

GLOBAL DTOthreshold, TwoPi AS EXT

GLOBAL UseDTOthreshold$

GLOBAL TotalFunctionCalls&

GLOBAL Quote$, UseFunction$, Use30Dschwefel$

GLOBAL ScreenWidth&, ScreenHeight&  'screen width & height

DECLARE SUB CFO(Nd%,Np%,Nt&,G,DeltaT,Alpha,Beta,Frep,R(),A(),M(),XiMin(),XiMax(),DiagLength,FunctionName$,LastStep&,Gamma,BestFitness,WorstFitness)

DECLARE FUNCTION SGO(R(),Nd%,p%,j&)         'SGO Function (2-D only)

DECLARE FUNCTION Schwefel226(R(),Nd%,p%,j&) 'SCHWEFEL Problem 2.26 (n-D)

DECLARE FUNCTION RASTRIGIN(R(),Nd%,p%,j&)   'RASTRIGIN (n-D)

DECLARE SUB NoProbesOnTheFloor(M(),R(),Nd%,ProbeNumber%,TimeStepNum&,FunctionName$,XiMin(),XiMax())

DECLARE SUB GetBestFitness(M(),Np%,StepNumber&,BestFitness,BestProbeNumber%,BestTimeStep&)

DECLARE SUB GetWorstFitness(M(),Np%,StepNumber&,WorstFitness,WorstProbeNumber%,WorstTimeStep&)

DECLARE SUB GetPlotAnnotation(PlotAnnotation$,Nd%,Np%,Nt&,G,DeltaT,Alpha,Beta,Frep,M(),FunctionName$,Gamma)

DECLARE SUB DisplayBestFitness(Np%,Nd%,LastStep&,M(),R(),BestFitnessProbeNumber%,BestFitnessTimeStep&,FunctionName$)

DECLARE SUB PlotBestFitnessEvolution(Nd%,Np%,LastStep&,G,DeltaT,Alpha,Beta,Frep,M(),FunctionName$,Gamma)

DECLARE SUB PlotAverageDistance(Nd%,Np%,LastStep&,G,DeltaT,Alpha,Beta,Frep,M(),FunctionName$,R(),DiagLength,Gamma)

DECLARE SUB InitialProbeDistribution(Np%,Nd%,Nt&,XiMin(),XiMax(),R(),Gamma)

DECLARE SUB RandomIPD(Np%,Nd%,Nt&,XiMin(),XiMax(),R())

DECLARE SUB TwoDplot(PlotFileName$,PlotTitle$,xCoord$,yCoord$,XaxisLabel$,YaxisLabel$, _
                     LogXaxis$,LogYaxis$,xMin$,xMax$,yMin$,yMax$,xTics$,yTics$,GnuPlotEXE$,LineType$,Annotation$)

DECLARE SUB ThreeDplot2(PlotFileName$,PlotTitle$,Annotation$,xCoord$,yCoord$,zCoord$, _
                        XaxisLabel$,YaxisLabel$,ZaxisLabel$,zMin$,zMax$,GnuPlotEXE$,A$,xStart$,xStop$,yStart$,yStop$)

DECLARE SUB Plot2Dfunction(XiMin(),XiMax(),Nd%,Np%,Nt&,PassNumber%,NumPasses%,FunctionName$)

DECLARE SUB CreateGNUplotINIfile(PlotWindowULC_X%,PlotWindowULC_Y%,PlotWindowWidth%,PlotWindowHeight%)

DECLARE SUB Delay(NumSecs)

DECLARE FUNCTION HasFITNESSsaturated$(Nsteps&,j&,Np%,Nd%,M(),R(),DiagLength)

DECLARE FUNCTION FormatFP$(X,Ndigits%)

DECLARE FUNCTION FormatInteger$(M%)

DECLARE FUNCTION UnitStep(X)

DECLARE FUNCTION RandomNum(a,b)

'======== MAIN PROGRAM =========

FUNCTION PBMAIN () AS LONG

'   ------ CFO Parameters -----

    LOCAL Nd%, Np%, Nt&

    LOCAL G, DeltaT, Alpha, Beta, Frep AS EXT

    LOCAL R(), A(), M() AS EXT 'position, acceleration & fitness matrices (usual CFO notation) NOTE THAT FITNESS M() IS NOT CFO 'MASS'!!
'                                                                                                                       ======
    LOCAL XiMin(), XiMax(), DiagLength AS EXT 'decision space boundaries, principal diagonal length

    LOCAL FunctionName$ 'name of objective function

'   ----------- Miscellaneous Setup Parameters --------------

    LOCAL Gamma, BestFitness, WorstFitness, WorstFitnessOverAllPasses, BestFitnessOverAllPasses, BestCoordinates(), RangeFraction, PreviousThreshold AS EXT

    LOCAL ThresholdValues(), BestFitnessesByPass() AS EXT

    LOCAL BestFitnessProbeNumber%, BestFitnessTimeStep&, GammaNum%, PassNumber%, NumPasses%, A$, B$

    LOCAL LastStep&, i%, N%

    TotalFunctionCalls& = 0

    RANDOMIZE TIMER 'seed random # gen with seconds//midnight (important in order to get different pseudorandom sequences on successive calls)

    Quote$ = CHR$(34) 'needed to set up plot files

    TwoPi = 8##*ATN(1##)

    DESKTOP GET SIZE TO ScreenWidth&, ScreenHeight&  'get screen size (global variables)

'   ------------------------------------ SCHWEFEL CHECK ------------------------------------
'    REDIM R(1 TO 1, 1 TO 2, 1 TO 1) :  R(1,1,1) = 420.9687## :  R(1,2,1) = 420.9687##
'    MSGBOX("Schwefel("+STR$(R(1,1,1))+","+STR$(R(1,2,1))+") ="+STR$(SCHWEFEL226(R(),2,1,1)))

    IF DIR$("wgnuplot.exe") = "" THEN MSGBOX("WARNING!  'wgnuplot.exe' not found.  Plots will not be displayed!") 'wgnuplot.exe is needed for on-screen plots

UseFunction$ = "SCHWEFEL226"  '"SGO" '"RASTRIGIN" '"SCHWEFEL226"  '"SGO" '"RASTRIGIN"

Use30Dschwefel$ = "YES" '"NO"

    IF UseFunction$   = "RASTRIGIN" THEN 'RASTRIGIN TEST FUNCTION (2-D)
        FunctionName$ = "RASTRIGIN" : Nd% = 2 : REDIM XiMin(1 TO Nd%), XiMax(1 TO Nd%) : FOR i% = 1 TO Nd% : XiMin(i%) = -5.12## : XiMax(i%) = 5.12## : NEXT i%
    END IF

    IF UseFunction$   = "SGO"       THEN 'SGO TEST FUNCTION (2-D)
        FunctionName$ = "SGO"       : Nd% = 2 : REDIM XiMin(1 TO Nd%), XiMax(1 TO Nd%) : FOR i% = 1 TO Nd% : XiMin(i%) = -50## : XiMax(i%) = 50## : NEXT i%
    END IF

    IF UseFunction$   = "SCHWEFEL226" THEN 'SCHWEFEL226 TEST FUNCTION (2-D)
        IF Use30Dschwefel$ = "YES" THEN
            FunctionName$ = "SCHWEFEL226" : Nd% = 30 : REDIM XiMin(1 TO Nd%), XiMax(1 TO Nd%) : FOR i% = 1 TO Nd% : XiMin(i%) = -500## : XiMax(i%) = 500## : NEXT i% '30-D
```

```
        ELSE
            FunctionName$ = "SCHWEFEL226" : Nd% = 2  : REDIM XiMin(1 TO Nd%), XiMax(1 TO Nd%) : FOR i% = 1 TO Nd% : XiMin(i%) = -500## : XiMax(i%) = 500## : NEXT i% '2-D
        END IF 'Use30Dschwefel$ = "YES"
    END IF

    DiagLength = 0## : FOR i% = 1 TO Nd% : DiagLength = DiagLength + (XiMax(i%)-XiMin(i%))^2 : NEXT i% : DiagLength = SQR(DiagLength)'LEN DS PRIN DIAG TO NORMALIZE Davg

'   ------------------------------------------------------- CFO LOOP -------------------------------------------------------

    G = 2## : DeltaT = 1## : Alpha = 2## : Beta = 2## : Frep = 0.5## : Gamma = 0.5## 'note: Gamma used only for Probe Line IPD

    REDIM BestCoordinates(1 TO Nd%) : BestFitnessOverAllPasses = -1E4200 : WorstFitnessOverAllPasses = 1E4200 : PreviousThreshold = -1E4200

    MinDistanceAboveThreshold = 0.005##

    NumPasses% = 6 : Nt& = 15 : Np% = 4 : UseDTOthreshold$ = "NO" : A$ = "NO Threshold."

    REDIM ThresholdValues(1 TO NumPasses%), BestFitnessesByPass(1 TO NumPasses%)

FOR PassNumber% = 1 TO NumPasses% 'SET NEW THRESHOLD ON EACH PASS (NO THRESHOLD ON INITIAL PASS)

    IF Nd% = 2 THEN CALL Plot2Dfunction(XiMin(),XiMax(),Nd%,Np%,Nt&,PassNumber%,NumPasses%,FunctionName$) 'plot 2-D function to see threshold

    FOR GammaNum% = 1 TO 11 'Gamma loop added 06-05-2012

        Gamma = (GammaNum%-1)/10##
        CALL CFO(Nd%,Np%,Nt&,G,DeltaT,Alpha,Beta,Frep,R(),A(),M(),XiMin(),XiMax(),DiagLength,FunctionName$,LastStep&,Gamma,BestFitness,WorstFitness)

        IF WorstFitness =< WorstFitnessOverAllPasses THEN WorstFitnessOverAllPasses = WorstFitness 'save worst overall fitness for setting DTO threshold

        IF PassNumber% > 1 THEN A$ = "Threshold = "+STR$(DTOthreshold)+"."

        IF Nd% = 2 THEN MSGBOX("Returned best fitness this run = "+STR$(BestFitness)+CHR$(13)+" with "+STR$(TotalFunctionCalls&)+ _
                              " total function calls using "+STR$(Np%)+" probes"+CHR$(13)+" with "+A$)

'       ------------------- DISPLAY BEST FITNESS, SAVE BEST OVERALL FITNESS, PLOT FITNESS/AVG DISTANCE EVOLUTION ---------------------

        CALL DisplayBestFitness(Np%,Nd%,LastStep&,M(),R(),BestFitnessProbeNumber%,BestFitnessTimeStep&,FunctionName$)

        BestFitnessesByPass(PassNumber%) = M(BestFitnessProbeNumber%,BestFitnessTimeStep&) 'save best fitnesses by pass #

        IF M(BestFitnessProbeNumber%,BestFitnessTimeStep&) >=  BestFitnessOverAllPasses THEN 'save best overall fitness & its coordinates
            BestFitnessOverAllPasses = M(BestFitnessProbeNumber%,BestFitnessTimeStep&)
            FOR i% = 1 TO Nd% : BestCoordinates(i%) = R(BestFitnessProbeNumber%,i%,BestFitnessTimeStep&) : NEXT i% 'copy coordinates
        END IF

        IF Nd% = 2 THEN
            CALL PlotBestFitnessEvolution(Nd%,Np%,LastStep&,G,DeltaT,Alpha,Beta,Frep,M(),FunctionName$,Gamma)
            CALL PlotAverageDistance(Nd%,Np%,LastStep&,G,DeltaT,Alpha,Beta,Frep,M(),FunctionName$,R(),DiagLength,Gamma)
        END IF

    NEXT GammaNum% 'added 06-05-2012

'   --------- Update Threshold & Number of Probes ---------

    ThresholdValues(PassNumber%) = DTOthreshold 'save threshold values by pass #

    RangeFraction = 0.60##*PassNumber%/NumPasses% 'changed 06-05-2012
'   RangeFraction = 0.98##*PassNumber%/NumPasses% 'commented out 06-05-2012
    DTOthreshold  = WorstFitnessOverAllPasses + RangeFraction*(BestFitnessOverAllPasses-WorstFitnessOverAllPasses)

    Np% = 2*Np% 'Increase # probes as DS landscape is flattened -> explore more aggressively because of likelihood that more probes are 'on the floor' (at threshold).

    UseDTOthreshold$ = "YES"

NEXT PassNumber%

    A$ = "RUN COMPLETED"+CHR$(13)+CHR$(13)+"Best Fitness Over All Passes = "+STR$(BestFitnessOverAllPasses)+CHR$(13)+"using"+STR$(TotalFunctionCalls&)+" function calls "+_
         "at coordinates "+CHR$(13)
    FOR i% = 1 TO Nd% : A$ = A$ + "x("+REMOVE$(STR$(i%),ANY" ")+") ="+STR$(BestCoordinates(i%))+CHR$(13) : NEXT i%

    B$ = "Pass#       Threshold              Best Fitness"+CHR$(13)
    FOR PassNumber% = 1 TO NumPasses%
        IF PassNumber% = 1 THEN
            B$ = B$ + USING$("  ##                  none              #######.#######",PassNumber%,BestFitnessesByPass(PassNumber%)) + CHR$(13)
        ELSE
            B$ = B$ + USING$("  ##              ########.###          #######.#######",PassNumber%,ThresholdValues(PassNumber%),BestFitnessesByPass(PassNumber%)) + CHR$(13)
        END IF
    NEXT PassNumber%

    MSGBOX(A$+CHR$(13)+B$)

    N% = FREEFILE : OPEN "DTO.DAT" FOR OUTPUT AS #N% : PRINT #N%,A$+$CRLF+B$ : CLOSE #N%

END FUNCTION 'PBMAIN(), MAIN PROGRAM

'=========================================================  CFO SUBROUTINE  ============================================================

SUB CFO(Nd%,Np%,Nt&,G,DeltaT,Alpha,Beta,Frep,R(),A(),M(),XiMin(),XiMax(),DiagLength,FunctionName$,LastStep&,Gamma,BestFitness,WorstFitness)

LOCAL p%, i%, j& 'Standard CFO indices: Probe# (1=<p%=<Np%), Coordinate# (1=<i%=<Nd%), Time Step# (0=<j&=<Nt&) <<!NOTE LOWER LIMIT ON j&!>>

LOCAL k%, L%     'Dummy summation indices

LOCAL SumSQ, Denom, Numerator AS EXT

LOCAL StepNumber&, BestProbeNumber%, BestTimeStep&, WorstProbeNumber%, WorstTimeStep&

LOCAL DavgOscillation$, DavgSaturation$, FitnessSaturation$

LOCAL NstepsDavgSat&, NstepsFitnessSat&

LOCAL RatioOfSuccessiveSlopes AS EXT

LOCAL StatusWindowHandle???

REDIM R(1 TO Np%, 1 TO Nd%, 0 TO Nt&), A(1 TO Np%, 1 TO Nd%, 0 TO Nt&), M(1 TO Np%, 0 TO Nt&) 'Re-initialize Position Vector/Acceleration/Fitness matrices

'STEP (A1) ----------------- Compute Initial Probe Distribution (Step 0)--------------------

'   CHANGED BACK TO PROBE LINE IPD 06-05-2012
    CALL InitialProbeDistribution(Np%,Nd%,Nt&,XiMin(),XiMax(),R(),Gamma) 'places probes uniformly on "probe lines" parallel to coordinate axes (see arXiv:1001.0317v1[cs.NE] at
www.arXiv.org for details)

'    CALL RandomIPD(Np%,Nd%,Nt&,XiMin(),XiMax(),R()) 'probes randomly placed throughout DS (avoids too many probes 'on the floor' -> no DS info) 'ADDED 05-29-2012

'STEP (A2) ------------- Compute Initial Fitness Matrix (Step 0) -------------

IF UseFunction$ = "RASTRIGIN" THEN
    FOR p% = 1 TO Np%
        M(p%,0) = RASTRIGIN(R(),Nd%,p%,0) : INCR TotalFunctionCalls&
        CALL NoProbesOnTheFloor(M(),R(),Nd%,p%,0,FunctionName$,XiMin(),XiMax()) 'THIS MAY NOT BE ESPECIALLY HELPFUL.  SEE SUBROUTINE BELOW...
    NEXT p%
END IF

IF UseFunction$ = "SGO" THEN
    FOR p% = 1 TO Np%
        M(p%,0) = SGO(R(),Nd%,p%,0) : INCR TotalFunctionCalls&
        CALL NoProbesOnTheFloor(M(),R(),Nd%,p%,0,FunctionName$,XiMin(),XiMax())
```

```
    NEXT p%
END IF

IF UseFunction$ = "SCHWEFEL226" THEN
    FOR p% = 1 TO Np%
        M(p%,0) = SCHWEFEL226(R(),Nd%,p%,0): INCR TotalFunctionCalls&
        CALL NoProbesOnTheFloor(M(),R(),Nd%,p%,0,FunctionName$,XiMin(),XiMax())
    NEXT p%
END IF

'STEP (A3) --------------------- Set Initial Probe Accelerations to Zero (Step 0)-------------------------

    FOR p% = 1 TO Np% : FOR i% = 1 TO Nd% : A(p%,i%,0) = 0## : NEXT i% : NEXT p% 'coord #, probe # (i%,p%)

'   =========================== LOOP ON TIME STEPS STARTING AT STEP #1 ================================

    LastStep& = Nt& 'unless run is terminated earlier

    FOR j& = 1 TO Nt&

'STEP (B) ----------- Compute Probe Position Vectors for this Time Step --------

        FOR p% = 1 TO Np% : FOR i% = 1 TO Nd% : R(p%,i%,j&) = R(p%,i%,j&-1) + 0.5##*A(p%,i%,j&-1)*DeltaT^2 : NEXT i% : NEXT p%

'STEP (C) ----------- Retrieve Errant Probes ----------- (see arXiv:1001.0317v1[cs.NE] at www.arXiv.org for deatils)

        FOR p% = 1 TO Np%

            FOR i% = 1 TO Nd%

                IF R(p%,i%,j&) < XiMin(i%) THEN R(p%,i%,j&) = XiMin(i%) + Frep*(R(p%,i%,j&-1)-XiMin(i%))

                IF R(p%,i%,j&) > XiMax(i%) THEN R(p%,i%,j&) = XiMax(i%) - Frep*(XiMax(i%)-R(p%,i%,j&-1))

            NEXT i%

        NEXT p%

'STEP (D) ----------- Compute Fitness Matrix for Current Probe Distribution ---------

IF UseFunction$ = "RASTRIGIN" THEN
       FOR p% = 1 TO Np%
           M(p%,j&) = RASTRIGIN(R(),Nd%,p%,j&): INCR TotalFunctionCalls&
           CALL NoProbesOnTheFloor(M(),R(),Nd%,p%,j&,FunctionName$,XiMin(),XiMax())
       NEXT p%
END IF

IF UseFunction$ = "SGO" THEN
       FOR p% = 1 TO Np%
           M(p%,j&) = SGO(R(),Nd%,p%,j&): INCR TotalFunctionCalls&
           CALL NoProbesOnTheFloor(M(),R(),Nd%,p%,j&,FunctionName$,XiMin(),XiMax())
       NEXT p%
END IF

IF UseFunction$ = "SCHWEFEL226" THEN
       FOR p% = 1 TO Np%
           M(p%,j&) = SCHWEFEL226(R(),Nd%,p%,j&): INCR TotalFunctionCalls&
           CALL NoProbesOnTheFloor(M(),R(),Nd%,p%,j&,FunctionName$,XiMin(),XiMax())
       NEXT p%
END IF

'STEP (E) ----------- Compute Accelerations Based on Current Probe Distribution & Fitnesses ----------------

        FOR p% = 1 TO Np%

            FOR i% = 1 TO Nd%

                A(p%,i%,j&) = 0

                FOR k% = 1 TO Np%

                    IF k% <> p% THEN

                        SumSQ = 0##

                        FOR L% = 1 TO Nd%  : SumSQ = SumSQ + (R(k%,L%,j&)-R(p%,L%,j&))^2 : NEXT L% 'dummy index

                        Denom = SQR(SumSQ) : Numerator = UnitStep((M(k%,j&)-M(p%,j&)))*(M(k%,j&)-M(p%,j&)) 'IMPORTANT NOTE: THIS CFO IMPLEMENTATION TAKES ADVANTAGE
                                                                                                        '==============
                                                                                                        'OF POWER BASIC'S AUTOMATIC HANDLING OF DIVIDE-BY-ZERO WHICH
                                                                                                        'AVOIDS "NaN" OR TERMINATION OF EXECUTION WHEN Denom = 0##.
                                                                                                        'MOST OTHER COMPILERS DO NOT CONTAIN THIS FEATURE AND
                                                                                                        'CONSEQUENTLY REQUIRE APPROPRIATE CODE MODIFICATIONS TO
                                                                                                        'AVOID A DIVIDE CHECK AND WHICH MAY RESULT IN SLIGHTLY
                                                                                                        'DIFFERENT RESULTS THAN THE BENCHMARK VALUES REPORTED HERE.
                                                                                                        'ADDED 05-28-2012.

                        A(p%,i%,j&) = A(p%,i%,j&) + G*(R(k%,i%,j&)-R(p%,i%,j&))*Numerator^Alpha/Denom^Beta

                    END IF

                NEXT k% 'dummy index

            NEXT i% 'coord (dimension) #

        NEXT p% 'probe #

'   --------- Get Best & Worst Fitnesses & Corresponding Probe #'s and Time Steps ---------

    CALL GetBestFitness(M(),Np%,j&,BestFitness,BestProbeNumber%,BestTimeStep&)

    CALL GetWorstFitness(M(),Np%,j&,WorstFitness,WorstProbeNumber%,WorstTimeStep&)

'    NstepsFitnessSat&  = 50 '# steps for averaging Fitness to test for saturation 'NOT USED

'     FitnessSaturation$ = HasFITNESSsaturated$(NstepsFitnessSat&,j&,Np%,Nd%,M(),R(),DiagLength) 'check for saturation of best fitness  'DITTO

'   ---- Frep is Variable => Adjust Value ---- See "Pseudorandomness in Central Force Optimization", arXiv:1001.0317v1[cs.NE] at www.arXiv.org for procedure details.

    Frep = Frep + 0.05##

    IF Frep > 1## THEN Frep = 0.05## 'keep Frep in range [0.05,1]

NEXT j& 'END OF TIME STEP LOOP

END SUB 'CFO()

'==========================================================================

SUB NoProbesOnTheFloor(M(),R(),Nd%,ProbeNumber%,TimeStepNum&,FunctionName$,XiMin(),XiMax())

LOCAL i%, k%, MaxIterations%

EXIT SUB 'Doesn't seem to help much (at least not for Schwefel)???  So don't do anything, but keep code... just in case...

    MaxIterations% = 10000

    IF UseDTOthreshold$ = "YES" AND M(ProbeNumber%,TimeStepNum&)-DTOthreshold < MinDistanceAboveThreshold THEN 'Probe is "On the Threshold Floor"
```

```
        k% = 0
        IF FunctionName$ = "RASTRGIN" THEN
            DO UNTIL M(ProbeNumber%,TimeStepNum&)-DTOthreshold >= MinDistanceAboveThreshold OR k% = MaxIterations%
                FOR i% = 1 TO Nd% : R(ProbeNumber%,i%,TimeStepNum&) = RandomNum(XiMin(i%),XiMax(i%)) : NEXT i% 'Randomly reposition probe until it's off the floor
                M(ProbeNumber%,TimeStepNum&) = RASTRIGIN(R(),Nd%,ProbeNumber%,TimeStepNum&) : INCR TotalFunctionCalls& : INCR k%
            LOOP
        END IF 'FunctionName$ = "RASTRGIN"
        IF FunctionName$ = "SGO" THEN
            DO UNTIL M(ProbeNumber%,TimeStepNum&)-DTOthreshold >= MinDistanceAboveThreshold OR k% = MaxIterations%
                FOR i% = 1 TO Nd% : R(ProbeNumber%,i%,TimeStepNum&) = RandomNum(XiMin(i%),XiMax(i%)) : NEXT i% 'Randomly reposition probe until it's off the floor
                M(ProbeNumber%,TimeStepNum&) = SGO(R(),Nd%,ProbeNumber%,TimeStepNum&) : INCR TotalFunctionCalls& : INCR k%
            LOOP
        END IF 'FunctionName$ = "SGO"
        IF FunctionName$ = "SCHWEFEL226" THEN
            DO UNTIL M(ProbeNumber%,TimeStepNum&)-DTOthreshold >= MinDistanceAboveThreshold OR k% = MaxIterations%
                FOR i% = 1 TO Nd% : R(ProbeNumber%,i%,TimeStepNum&) = RandomNum(XiMin(i%),XiMax(i%)) : NEXT i% 'Randomly reposition probe until it's off the floor
                M(ProbeNumber%,TimeStepNum&) = SCHWEFEL226(R(),Nd%,ProbeNumber%,TimeStepNum&) : INCR TotalFunctionCalls& : INCR k%
            LOOP
        END IF 'FunctionName$ = "SCHWEFEL226"
    END IF 'UseDTOthreshold$ = "YES"
END SUB 'NoProbesOnTheFloor()
'----------------------------
SUB GetBestFitness(M(),Np%,StepNumber&,BestFitness,BestProbeNumber%,BestTimeStep&)
LOCAL p%, i&, A$
    BestFitness = M(1,0)
    FOR i& = 0 TO StepNumber&
        FOR p% = 1 TO Np%
            IF M(p%,i&) >= BestFitness THEN
                BestFitness = M(p%,i&) : BestProbeNumber% = p% : BestTimeStep& = i&
            END IF
        NEXT p%
    NEXT i&
END SUB 'GetBestFitness()
'------------------------
SUB GetWorstFitness(M(),Np%,StepNumber&,WorstFitness,WorstProbeNumber%,WorstTimeStep&)
LOCAL p%, i&, A$
    WorstFitness = M(1,0)
    FOR i& = 0 TO StepNumber&
        FOR p% = 1 TO Np%
            IF M(p%,i&) =< WorstFitness THEN
                WorstFitness = M(p%,i&) : WorstProbeNumber% = p% : WorstTimeStep& = i&
            END IF
        NEXT p%
    NEXT i&
END SUB 'GetWorstFitness()
'------------------------
FUNCTION SGO(R(),Nd%,p%,j&) 'SGO Function (2-D, cannot change dimensionality!)
'Reference:
'----------
'Hsiao, Y., Chuang, C., Jiang, J., and Chien, C., "A Novel Optimization Algorithm: Space
'Gravitational Optimization," Proc. of 2005 IEEE International Conference on Systems, Man,
'and Cybernetics, 3, 2323-2328. (2005)
'KNOWN MAXIMUM = ~130.8323226... @ ~(-2.8362075...,-2.8362075...) WITH ZERO OFFSET.
LOCAL x1, x2, Z, U, t1, t2, SGOx1offset, SGOx2offset AS EXT
    SGOx1offset = 0## : SGOx2offset = 0##
    x1 = R(p%,1,j&) - SGOx1offset : x2 = R(p%,2,j&) - SGOx2offset
    t1 = x1^4 - 16##*x1^2 + 0.5##*x1 : t2 = x2^4 - 16##*x2^2 + 0.5##*x2
    Z = t1 + t2
    U = -Z 'added 05/29/2012
    IF UseDTOthreshold$ = "YES" THEN U = (U-DTOthreshold)*UnitStep(U-DTOthreshold)+DTOthreshold 'added 05/29/2012 to include DTO threshold
    SGO = U
END FUNCTION 'SGO()
'------------------
FUNCTION SCHWEFEL226(R(),Nd%,p%,j&) 'Schwefel Problem 2.26 (n-D)
'MAX = 418.9829*Nd @ [420.9687]^Nd FOR ARBITRARY DIMENSIONALITY.
'MAX = 12,569.487 @ [420.9687]^30 (30-D CASE).
'MAX = 837.9658 @ (420.9687,420.9687) (2-D CASE).
```

```
'Reference:
'----------
'Yao, X., Liu, Y., and Lin, G., "Evolutionary Programming Made Faster,"
'IEEE Trans. Evolutionary Computation, Vol. 3, No. 2, 82-102, July 1999.

    LOCAL Z, U, Xi AS EXT

    LOCAL i%

    Z = 0##

    FOR i% = 1 TO Nd%

        Xi = R(p%,i%,j&)

        Z = Z + Xi*SIN(SQR(ABS(Xi)))

    NEXT i%

    U = Z 'added 05/29/2012

    IF UseDTOthreshold$ = "YES" THEN U = (U-DTOthreshold)*UnitStep(U-DTOthreshold)+DTOthreshold 'added 05/29/2012 to include DTO threshold

    Schwefel226 = U

END FUNCTION 'SCHWEFEL226()

'--------------------------

FUNCTION RASTRIGIN(R(),Nd%,p%,j&) '(n-D) RASTRIGIN  'MOD 02/29/2012 TO OFFSET MAX VAKUE & LOCATION

'MAXIMUM = ZERO (n-D CASE). DOMAIN IS [-5.12,5.12]^n.

'Reference:
'----------
'Yao, X., Liu, Y., and Lin, G., "Evolutionary Programming Made Faster,"
'IEEE Trans. Evolutionary Computation, Vol. 3, No. 2, 82-102, Jul. 1999.

    LOCAL Z, Xi, x1offset, x2offset, U AS EXT

    LOCAL i%

    x1offset = -1.25## : x2offset = 3.25##  'move max to (-1.25,3.25)

    Z = 0##

    FOR i% = 1 TO Nd%

        Xi = R(p%,i%,j&)

        IF Nd% = 2 THEN
            SELECT CASE i%
                CASE 1 : Xi = Xi - x1offset
                CASE 2 : Xi = Xi - x2offset
            END SELECT
        END IF

        Z  = Z + (Xi^2 - 10##*COS(TwoPi*Xi) + 10##)^2

    NEXT i%

    U = -Z + 10.123## 'increase max from zero to 10.123

    IF UseDTOthreshold$ = "YES" THEN U = (U-DTOthreshold)*UnitStep(U-DTOthreshold)+DTOthreshold 'added 05/29/2012 to include DTO threshold

    RASTRIGIN = U

END FUNCTION

'-----------

SUB RandomIPD(Np%,Nd%,Nt&,XiMin(),XiMax(),R()) 'RANDOM IPD USED FOR DTO

LOCAL p%, i%

    FOR p% = 1 TO Np% 'probe-by-probe

        FOR i% = 1 TO Nd%

            R(p%,i%,0) = RandomNum(XiMin(i%),XiMax(i%)) 'set each coord to random value between min and max

        NEXT i%

    NEXT p%

END SUB 'RandomIPD()

'-------------------

SUB InitialProbeDistribution(Np%,Nd%,Nt&,XiMin(),XiMax(),R(),Gamma) 'PROBE LINE IPD

'Places probes along "probe lines" parallel to cordinate axes with intersection point
'on decision space principal diagonal whose location is determined by parameter Gamma.
'See "Pseudorandomness in Central Force Optimization", arXiv:1001.0317v1[cs.NE] at www.arXiv.org.

LOCAL DeltaXi, DelX1, DelX2, Di AS EXT

LOCAL NumProbesPerAxis%, p%, i%, k%, NumX1points%, NumX2points%, x1pointNum%, x2pointNum%, A$

    NumProbesPerAxis% = Np%\Nd% 'even #

    FOR i% = 1 TO Nd%

        FOR p% = 1 TO Np%

            R(p%,i%,0) = XiMin(i%) + Gamma*(XiMax(i%)-XiMin(i%))

        NEXT Np%

    NEXT i%

    FOR i% = 1 TO Nd% 'place probes axis-by-axis (i% is axis [dimension] number)

        DeltaXi = (XiMax(i%)-XiMin(i%))/(NumProbesPerAxis%-1)

        FOR k% = 1 TO NumProbesPerAxis%

            p% = k% + NumProbesPerAxis%*(i%-1) 'probe #

            R(p%,i%,0) = XiMin(i%) + (k%-1)*DeltaXi

        NEXT k%

    NEXT i%

END SUB 'InitialProbeDistribution()

'----------------------------------
```

```
FUNCTION HasFITNESSsaturated$(Nsteps&,j&,Np%,Nd%,M(),R(),DiagLength) 'NOT USED
LOCAL A$
LOCAL k&, p%
LOCAL BestFitness, SumOfBestFitnesses, BestFitnessStepJ, FitnessSatTOL AS EXT
    A$ = "NO"
    FitnessSatTOL = 0.000001## 'tolerance for FITNESS saturation
    IF j& < Nsteps& + 10 THEN GOTO ExitHasFITNESSsaturated 'execute at least 10 steps after averaging interval before performing this check
    SumOfBestFitnesses = 0##
    FOR k& = j&-Nsteps&+1 TO j&
        BestFitness = M(k&,1)
        FOR p% = 1 TO Np%
            IF M(p%,k&) >= BestFitness THEN BestFitness = M(p%,k&)
        NEXT p%
        IF k& = j& THEN BestFitnessStepJ = BestFitness
        SumOfBestFitnesses = SumOfBestFitnesses + BestFitness
    NEXT k&
    IF ABS(SumOfBestFitnesses/Nsteps&-BestFitnessStepJ) =< FitnessSatTOL THEN A$ = "YES" 'saturation if (avg value - last value) are within TOL
ExitHasFITNESSsaturated:
    HasFITNESSsaturated$ = A$
END FUNCTION 'HasFITNESSsaturated$()
'-----------------------------------
SUB PlotBestFitnessEvolution(Nd%,Np%,LastStep&,G,DeltaT,Alpha,Beta,Frep,M(),FunctionName$,Gamma)
LOCAL BestFitness(), GlobalBestFitness AS EXT
LOCAL PlotAnnotation$, PlotTitle$
LOCAL p%, j&, N%
    REDIM BestFitness(0 TO LastStep&)
    CALL GetPlotAnnotation(PlotAnnotation$,Nd%,Np%,LastStep&,G,DeltaT,Alpha,Beta,Frep,M(),FunctionName$,Gamma)
    GlobalBestFitness = M(1,0)
    FOR j& = 0 TO LastStep&
        BestFitness(j&) = M(1,j&)
        FOR p% = 1 TO Np%
            IF M(p%,j&) >= BestFitness(j&)   THEN BestFitness(j&)   = M(p%,j&)
            IF M(p%,j&) >= GlobalBestFitness THEN GlobalBestFitness = M(p%,j&)
        NEXT p% 'probe #
    NEXT j& 'time step
    N% = FREEFILE
    OPEN "Fitness" FOR OUTPUT AS #N%
        FOR j& = 0 TO LastStep& : PRINT #N%, USING$("###### #######.#######",j&,BestFitness(j&)) : NEXT j&
    CLOSE #N%
    PlotAnnotation$ = PlotAnnotation$ + "Best Fitness = " + REMOVE$(STR$(ROUND(GlobalBestFitness,8)),ANY" ")
    PlotTitle$ = "Best Fitness vs Time Step\n" + "[" + REMOVE$(STR$(Np%),ANY" ") + " probes, "+REMOVE$(STR$(LastStep&),ANY" ")+" time steps]"
    CALL CreateGNUplotINIfile(0.1##*ScreenWidth&,0.1##*ScreenHeight&,0.6##*ScreenWidth&,0.6##*ScreenHeight&)
    CALL TwoDplot("Fitness","Best Fitness","0.7","0.7","Time Step\n\n.",".\n\nBest Fitness(X)", _
                  "","","","","","","","","wgnuplot.exe"," with lines linewidth 2",PlotAnnotation$)
END SUB 'PlotBestFitnessEvolution()
'----------------------------------
SUB PlotAverageDistance(Nd%,Np%,LastStep&,G,DeltaT,Alpha,Beta,Frep,M(),FunctionName$,R(),DiagLength,Gamma)
LOCAL Davg(), BestFitness(), TotalDistanceAllProbes, SumSQ AS EXT
LOCAL PlotAnnotation$, PlotTitle$
LOCAL p%, j&, N%, i%, BestProbeNumber%(), BestTimeStep&()
    REDIM Davg(0 TO LastStep&), BestFitness(0 TO LastStep&), BestProbeNumber%(0 TO LastStep&), BestTimeStep&(0 TO LastStep&)
    CALL GetPlotAnnotation(PlotAnnotation$,Nd%,Np%,LastStep&,G,DeltaT,Alpha,Beta,Frep,M(),FunctionName$,Gamma)
'   ----------- Best Probe #, etc. -----------
    FOR j& = 0 TO LastStep&
        BestFitness(j&) = M(1,j&)
        FOR p% = 1 TO Np%
            IF M(p%,j&) >= BestFitness(j&) THEN
                BestFitness(j&) = M(p%,j&) : BestProbeNumber%(j&) = p% : BestTimeStep&(j&) = j& 'only probe number is used at this time, but other data are computed for possible
future use.
            END IF
        NEXT p% 'probe #
    NEXT j& 'time step
    N% = FREEFILE
'   --------- Average Distance to Best Probe -----------
    FOR j& = 0 TO LastStep&
```

```
        TotalDistanceAllProbes = 0##
        FOR p% = 1 TO Np%
            SumSQ = 0##
            FOR i% = 1 TO Nd%
                SumSQ = SumSQ + (R(BestProbeNumber%(j&),i%,j&)-R(p%,i%,j&))^2 'do not exclude p%=BestProbeNumber%(j&) from sum because it adds zero
            NEXT i%
            TotalDistanceAllProbes = TotalDistanceAllProbes + SQR(SumSQ)
        NEXT p%
        Davg(j&) = TotalDistanceAllProbes/(DiagLength*(Np%-1)) 'but exclude best prove from average
    NEXT j&
'   ----------- Create Plot Data File -----------
    OPEN "Davg" FOR OUTPUT AS #N%
        FOR j& = 0 TO LastStep&
            PRINT #N%, USING$("###### #######.#######",j&,Davg(j&))
        NEXT j&
    CLOSE #N%
    PlotTitle$ = "Average Distance of " + REMOVE$(STR$(Np%-1),ANY" ") + " Probes to Best Probe\nNormalized to Size of Decision Space\n" + _
                 "[" + REMOVE$(STR$(Np%),ANY" ") + " probes, " + REMOVE$(STR$(LastStep&),ANY" ") + " time steps]"
    CALL CreateGNUplotINIfile(0.2##*ScreenWidth&,0.2##*ScreenHeight&,0.6##*ScreenWidth&,0.6##*ScreenHeight&)
    CALL TwoDplot("Davg",PlotTitle$,"0.7","0.9","Time Step\n\n.",".\n\n<D>/Ldiag", _
                  "","","","","","","","wgnuplot.exe"," with lines linewidth 2",PlotAnnotation$)
END SUB 'PlotAverageDistance()
'-----------------------------
SUB GetPlotAnnotation(PlotAnnotation$,Nd%,Np%,LastStep&,G,DeltaT,Alpha,Beta,Frep,M(),FunctionName$,Gamma)
LOCAL A$
    A$ = ""
    PlotAnnotation$ = FunctionName$ + " Function" + " ("+ FormatInteger$(Nd%) + "-D) \n"    +_
                      FormatInteger$(Np%) + " probes"        + A$ + "\n" +_
                      "G = " + FormatFP$(G,2)               + "\n" +_
                      "Alpha = "     + FormatFP$(Alpha,1)    + "\n" +_
                      "Beta = "      + FormatFP$(Beta,1)     + "\n" +_
                      "DelT = "      + FormatFP$(DeltaT,1)  + "\n"' + "Frep = "      + FormatFP$(Frep,3)    + "\n" + "Gamma = "     + FormatFP$(Gamma,3)   + "\n"
END SUB 'GetPlotAnnotation()
'---------------------------
SUB DisplayBestFitness(Np%,Nd%,LastStep&,M(),R(),BestFitnessProbeNumber%,BestFitnessTimeStep&,FunctionName$)
LOCAL A$, B$, p%, i%, j&
LOCAL BestFitness AS EXT
    B$ = "" : IF Nd% > 1 THEN B$ = "s"
    BestFitness = M(1,0)
    FOR j& = 0 TO LastStep&
        FOR p% = 1 TO Np%
            IF M(p%,j&) >= BestFitness THEN
                BestFitness = M(p%,j&) : BestFitnessProbeNumber% = p% : BestFitnessTimeStep& = j&
            END IF
        NEXT p%
    NEXT j&
    A$ = FunctionName$ + CHR$(13) +_
         "Best Fitness = " + REMOVE$(STR$(ROUND(BestFitness,8)),ANY" ") + " returned by" + CHR$(13)        +_
         "Probe # "        + REMOVE$(STR$(BestFitnessProbeNumber%),ANY" ") +_
         " at Time Step "   + REMOVE$(STR$(BestFitnessTimeStep&),ANY" ") + CHR$(13) + CHR$(13) + "P" + REMOVE$(STR$(BestFitnessProbeNumber%),ANY" ") + " coordinate" + B$ + ":" +
CHR$(13)
    FOR i% = 1 TO Nd% : A$ = A$ + STR$(i%)+"    "+REMOVE$(STR$(ROUND(R(BestFitnessProbeNumber%,i%,BestFitnessTimeStep&),8)),ANY" ")+CHR$(13) : NEXT i%
    IF Nd% = 2 THEN MSGBOX(A$) 'mod 06-05-2012
END SUB 'DisplayBestFitness()
'----------------------------
FUNCTION FormatInteger$(M%) : FormatInteger$ = REMOVE$(STR$(M%),ANY" ") : END FUNCTION 'formats integer variable for display
'----------------------------
FUNCTION FormatFP$(X,Ndigits%) 'formats floating point variable for display
LOCAL A$
    IF X = 0## THEN
        A$ = "0." : GOTO ExitFormatFP
    END IF
    A$ = REMOVE$(STR$(ROUND(ABS(X),Ndigits%)),ANY" ")
    IF ABS(X) < 1## THEN
        IF X > 0## THEN
            A$ = "0" + A$
        ELSE
            A$ = "-0" + A$
        END IF
    ELSE
```

```
        IF X < 0## THEN A$ = "-" + A$
    END IF
ExitFormatFP:
    FormatFP$ = A$
END FUNCTION 'FormatFP$()
'------------------------
FUNCTION UnitStep(X)
LOCAL Z AS EXT
    IF X < 0## THEN
        Z = 0##
    ELSE
        Z = 1##
    END IF
    UnitStep = Z
END FUNCTION 'UnitStep()
'-----------------------
FUNCTION RandomNum(a,b) 'Returns random number X where a =< X < b.
    RandomNum = a + (b-a)*RND
END FUNCTION 'RandomNum()
'------------------------
    SUB TwoDplot(PlotFileName$,PlotTitle$,xCoord$,yCoord$,XaxisLabel$,YaxisLabel$, _
                 LogXaxis$,LogYaxis$,xMin$,xMax$,yMin$,yMax$,xTics$,yTics$,GnuPlotEXE$,LineType$,Annotation$)
        LOCAL N%, ProcessID???
        N% = FREEFILE
        OPEN "cmd2d.gp" FOR OUTPUT AS #N%
            IF LogXaxis$ = "YES" AND LogYaxis$ = "NO"  THEN PRINT #N%, "set logscale x"
            IF LogXaxis$ = "NO"  AND LogYaxis$ = "YES" THEN PRINT #N%, "set logscale y"
            IF LogXaxis$ = "YES" AND LogYaxis$ = "YES" THEN PRINT #N%, "set logscale xy"
            IF xMin$ <> "" AND xMax$ <> "" THEN  PRINT #N%, "set xrange ["+xMin$+":"+xMax$+"]"
            IF yMin$ <> "" AND yMax$ <> "" THEN  PRINT #N%, "set yrange ["+yMin$+":"+yMax$+"]"
            PRINT #N%, "set label "      + Quote$ + Annotation$ + Quote$ + " at graph " + xCoord$ + "," + yCoord$
            PRINT #N%, "set grid xtics " + XTics$
            PRINT #N%, "set grid ytics " + yTics$
            PRINT #N%, "set grid mxtics"
            PRINT #N%, "set grid mytics"
            PRINT #N%, "show grid"
            PRINT #N%, "set title "  + Quote$+PlotTitle$+Quote$
            PRINT #N%, "set xlabel " + Quote$+XaxisLabel$+Quote$
            PRINT #N%, "set ylabel " + Quote$+YaxisLabel$+Quote$
            PRINT #N%, "plot "+Quote$+PlotFileName$+Quote$+" notitle"+LineType$
        CLOSE #N%
        ProcessID??? = SHELL(GnuPlotEXE$+" cmd2d.gp -") : CALL Delay(0.5##) 'to give OS time...
    END SUB 'TwoDplot()
'----------------------
    SUB CreateGNUplotINIfile(PlotWindowULC_X%,PlotWindowULC_Y%,PlotWindowWidth%,PlotWindowHeight%)
    LOCAL N%, WinPath$, A$, B$, WindowsDirectory$
    WinPath$ = UCASE$(ENVIRON$("Path"))'DIR$("C:\WINDOWS",23)
    DO
        B$ = A$
        A$ = EXTRACT$(WinPath$,";")
        WinPath$ = REMOVE$(WinPath$,A$+";")
        IF RIGHT$(A$,7) = "WINDOWS" OR A$ = B$ THEN EXIT LOOP
        IF RIGHT$(A$,5) = "WINNT"   OR A$ = B$ THEN EXIT LOOP
    LOOP
    WindowsDirectory$ = A$
    N% = FREEFILE
'   ----------- WGNUPLOT.INPUT FILE -----------
    OPEN WindowsDirectory$+"\wgnuplot.ini" FOR OUTPUT AS #N%
        PRINT #N%,"[WGNUPLOT]"
        PRINT #N%,"TextOrigin=0 0"
        PRINT #N%,"TextSize=640 150"
        PRINT #N%,"TextFont=Terminal,9"
        PRINT #N%,"GraphOrigin="+REMOVE$(STR$(PlotWindowULC_X%),ANY" ")+" "+REMOVE$(STR$(PlotWindowULC_Y%),ANY" ")
        PRINT #N%,"GraphSize="  +REMOVE$(STR$(PlotWindowWidth%),ANY" ")+" "+REMOVE$(STR$(PlotWindowHeight%),ANY" ")
        PRINT #N%,"GraphFont=Arial,10"
        PRINT #N%,"GraphColor=1"
        PRINT #N%,"GraphToTop=1"
        PRINT #N%,"GraphBackground=255 255 255"
        PRINT #N%,"Border=0 0 0 0 0"
        PRINT #N%,"Axis=192 192 192 2 2"
        PRINT #N%,"Line1=0 0 255 0 0"
        PRINT #N%,"Line2=0 255 0 0 1"
        PRINT #N%,"Line3=255 0 0 0 2"
        PRINT #N%,"Line4=255 0 255 0 3"
        PRINT #N%,"Line5=0 0 128 0 4"
    CLOSE #N%
    END SUB 'CreateGNUplotINIfile()
'----------------------------------
    SUB Delay(NumSecs)
        LOCAL StartTime, StopTime AS EXT
```

```
        StartTime = TIMER

        DO UNTIL (StopTime-StartTime) >= NumSecs

            StopTime = TIMER

        LOOP

    END SUB 'Delay()

'-------------------

    SUB ThreeDplot2(PlotFileName$,PlotTitle$,Annotation$,xCoord$,yCoord$,zCoord$, _
                XaxisLabel$,YaxisLabel$,ZaxisLabel$,zMin$,zMax$,GnuPlotEXE$,A$,xStart$,xStop$,yStart$,yStop$)
        LOCAL N%

        N% = FREEFILE

        OPEN "cmd3d.gp" FOR OUTPUT AS #N%

            PRINT #N%, "set pm3d"
            PRINT #N%, "show pm3d"
            PRINT #N%, "set hidden3d"
            PRINT #N%, "set view 45, 45, 1, 1"

            IF zMin$ <> "" AND zMax$ <> "" THEN  PRINT #N%, "set zrange ["+zMin$+":"+zMax$+"]"

            PRINT #N%, "set xrange [" + xStart$ + ":" + xStop$ + "]"
            PRINT #N%, "set yrange [" + yStart$ + ":" + yStop$ + "]"

            PRINT #N%, "set label "   + Quote$  + AnnoTation$ + Quote$+" at graph "+xCoord$+","+yCoord$+","+zCoord$
            PRINT #N%, "show label"
            PRINT #N%, "set grid xtics ytics ztics"
            PRINT #N%, "show grid"
            PRINT #N%, "set title "  + Quote$+PlotTitle$    + Quote$
            PRINT #N%, "set xlabel " + Quote$+XaxisLabel$   + Quote$
            PRINT #N%, "set ylabel " + Quote$+YaxisLabel$   + Quote$
            PRINT #N%, "set zlabel " + Quote$+ZaxisLabel$   + Quote$
            PRINT #N%, "splot "      + Quote$+PlotFileName$ + Quote$ + A$ + " notitle with lines"
        CLOSE #N%

        SHELL(GnuPlotEXE$+" cmd3d.gp -")

    END SUB 'ThreeDplot2()

'-------------------------

SUB Plot2Dfunction(XiMin(),XiMax(),Nd%,Np%,Nt&,PassNumber%,NumPasses%,FunctionName$)

LOCAL NumPoints%, i%, k%, N%, A$, B$

LOCAL DelX1, DelX2, Z, R() AS EXT

    REDIM R(1 TO Np%, 1 TO Nd%, 0 TO Nt&)

    NumPoints% = 100

    N% = FREEFILE : OPEN "TwoDplot.DAT" FOR OUTPUT AS #N%

    DelX1 = (XiMax(1)-XiMin(1))/(NumPoints%-1) : DelX2 = (XiMax(2)-XiMin(2))/(NumPoints%-1)

    FOR i% = 1 TO NumPoints%

        R(1,1,0) = XiMin(1) + (i%-1)*DelX1 'x1 value

        FOR k% = 1 TO NumPoints%

            R(1,2,0) = XiMin(2) + (k%-1)*DelX2 'x2 value

            IF FunctionName$ = "SGO"         THEN Z = SGO(R(),2,1,0)

            IF FunctionName$ = "RASTRIGIN"   THEN Z = RASTRIGIN(R(),2,1,0)

            IF FunctionName$ = "SCHWEFEL226" THEN Z = SCHWEFEL226(R(),2,1,0)

            PRINT #N%, USING$("######.###### ######.###### #######.######^^^^",R(1,1,0),R(1,2,0),Z)

        NEXT k%

        PRINT #N%, ""

    NEXT i%

    CLOSE #N%

    REDIM R(1 TO Np%, 1 TO Nd%, 0 TO Nt&)

    CALL CreateGNUplotINIfile(0.1##*ScreenWidth&,0.1##*ScreenHeight&,0.6##*ScreenWidth&,0.6##*ScreenHeight&)

    A$ = ", No Threshold" : IF PassNumber% > 1 THEN A$ = ", Threshold = "+REMOVE$(STR$(ROUND(DTOthreshold,2)),ANY" ")

    B$ = ""
    IF FunctionName$ = "RASTRIGIN"   THEN B$ = "\n Known max = 10.123 @ (-1.25,3.25)\n"
    IF FunctionName$ = "SGO"         THEN B$ = "\n Known max ~130.8323226... @ ~(-2.8362075...,-2.8362075...)\n"
    IF FunctionName$ = "SCHWEFEL226" THEN B$ = "\n Known max 837.9658 @ (420.9687,420.9687)\n"

    CALL ThreeDplot2("TwoDplot.DAT",FunctionName$+", Pass #"+REMOVE$(STR$(PassNumber%),ANY" ")+"/"+REMOVE$(STR$(NumPasses%),ANY" ")+A$+B$,_
                 "","0.6","0.6","1.2","x1","x2","z=f(x1,x2)","","","wgnuplot.exe","","","","","")

END SUB 'Plot2Dfunction()

'***************************************************  END PROGRAM 'DTO_TEST_06-05-2012.BAS'   ***********************************************
```